\documentclass[fleqn,usenatbib]{mnras}

\usepackage{newtxtext,newtxmath}

\usepackage[T1]{fontenc}
\usepackage{ae,aecompl}

\usepackage{graphicx}	
\usepackage{amsmath}	
\usepackage{amssymb}	
\usepackage{subfig}
\usepackage[dvipsnames]{xcolor}
\usepackage[normalem]{ulem}
\usepackage{times}

\makeatletter\g@addto@macro{\UrlBreaks}{\UrlOrds}\makeatother

\usepackage{comment}

\newcommand{\AR}[1]{\textcolor{cyan}{(AR: #1)}}

\newcommand{\vect}[1]{\boldsymbol{#1}}
\newcommand{\mpc}{$\, h^{-1}$Mpc }

\newcommand\Mpch{\, h^{-1}\mathrm{Mpc}}

\newcommand\spara{s_\parallel}
\newcommand\sperp{s_\perp}
\newcommand\apara{a_\parallel}
\newcommand\aperp{a_\perp}
\newcommand\sig{\sigma_{8}} 
\newcommand\fsig{f\sig}

\title[ELG eBOSS RSD measurements]
{The Completed SDSS-IV extended Baryon Oscillation Spectroscopic Survey: Growth rate of structure measurement from anisotropic clustering analysis in configuration space between redshift 0.6 and 1.1 for the Emission Line Galaxy sample}
\author[A. Tamone et al.]{\parbox{\textwidth}{
Am\'elie Tamone$^{1\thanks{e-mail: \texttt{\href{E-mail:amelie.tamone@epfl.ch}{amelie.tamone@epfl.ch}}}}$,
Anand Raichoor$^{1}$,
Cheng Zhao$^{1}$,
Arnaud de Mattia$^{2}$,
Claudio Gorgoni$^{1}$,
Etienne Burtin$^{2}$,
Vanina Ruhlmann-Kleider$^{2}$,
Ashley J. Ross$^{3}$,
Shadab Alam$^{4}$,
Will J. Percival$^{5,6,7}$,
Santiago Avila$^{8,9}$,
Michael J. Chapman$^{5,6}$,
Chia-Hsun Chuang$^{10}$,
Johan Comparat$^{11}$,
Kyle S. Dawson$^{12}$,
Sylvain de la Torre$^{13}$,
H\'elion du Mas des Bourboux$^{12}$,
Stephanie Escoffier$^{13}$,
Violeta Gonzalez-Perez$^{14,15}$,
Jiamin Hou$^{11}$,
Jean-Paul Kneib$^{1,13}$,
Faizan G. Mohammad$^{5,6}$,
Eva-Maria Mueller$^{16}$,
Romain Paviot$^{13}$,
Graziano Rossi$^{17}$,
Donald P. Schneider$^{18,19}$,
Yuting Wang$^{20}$,
Gong-Bo Zhao$^{20,21,15}$
 } \vspace*{4pt} \\ 
\small $^{1}$ Institute of Physics, Laboratory of Astrophysics, Ecole Polytechnique F\'ed\'erale de Lausanne (EPFL), Observatoire de Sauverny, 1290 Versoix, Switzerland\vspace*{-2pt} \\ 
\small $^{2}$ IRFU, CEA, Universit\'e Paris-Saclay, F-91191 Gif-sur-Yvette, France\vspace*{-2pt} \\ 
\small $^{3}$ Center for Cosmology and AstroParticle Physics, The Ohio State University, Columbus, OH 43212\vspace*{-2pt} \\ 
\small $^{4}$ Institute for Astronomy, University of Edinburgh, Royal Observatory, Blackford Hill, Edinburgh, EH9 3HJ , UK\vspace*{-2pt} \\ 
\small $^{5}$ Waterloo Centre for Astrophysics, University of Waterloo, Waterloo, ON~N2L~3G1, Canada\vspace*{-2pt} \\ 
\small $^{6}$ Department of Physics and Astronomy, University of Waterloo, 200 University Ave W, Waterloo, ON N2L 3G1, Canada\vspace*{-2pt} \\ 
\small $^{7}$ Perimeter Institute, Waterloo, ON~N2L~2Y5, Canada\vspace*{-2pt} \\ 
\small $^{8}$ Universidad Aut\'onoma de Madrid, 28049, Madrid, Spain\vspace*{-2pt} \\ 
\small $^{9}$ Instituto de Fisica Teorica UAM/CSIC, Universidad Autonoma de Madrid, 28049 Madrid, Spain\vspace*{-2pt} \\ 
\small $^{10}$ Kavli Institute for Particle Astrophysics and Cosmology, Stanford University, 452 Lomita Mall, Stanford, CA 94305, USA\vspace*{-2pt} \\ 
\small $^{11}$ Max-Planck-Institut f\"{u}r extraterrestrische Physik (MPE), Giessenbachstrasse 1, D-85748 Garching bei M\"unchen, Germany\vspace*{-2pt} \\ 
\small $^{12}$ University of Utah, Department of Physics and Astronomy, 115 S 1400 E, Salt Lake City, UT 84112, USA\vspace*{-2pt} \\ 
\small $^{13}$ Aix Marseille Univ, CNRS, CNES, LAM, Marseille, France\vspace*{-2pt} \\ 
\small $^{14}$ Astrophysics Research Institute, Liverpool John Moores University, 146 Brownlow Hill, Liverpool L3 5RF, UK\vspace*{-2pt} \\ 
\small $^{15}$ Institute of Cosmology \& Gravitation, University of Portsmouth, Dennis Sciama Building, Burnaby Road, Portsmouth PO1 3FX, UK\vspace*{-2pt} \\ 
\small $^{16}$ Department of Physics, University of Oxford, Denys Wilkinson Building, Keble Road, Oxford OX1 3RH, U.K\vspace*{-2pt} \\ 
\small $^{17}$ Department of Physics and Astronomy, Sejong University, Seoul, 143-747, Korea\vspace*{-2pt} \\ 
\small $^{18}$ Department of Astronomy and Astrophysics, The Pennsylvania State University, University Park, PA 16802\vspace*{-2pt} \\ 
\small $^{19}$ Institute for Gravitation and the Cosmos, The Pennsylvania State University, University Park, PA 16802\vspace*{-2pt} \\ 
\small $^{20}$ National Astronomical Observatories of China, Chinese Academy of Sciences, 20A Datun Road, Chaoyang District, Beijing 100012, China\vspace*{-2pt} \\ 
\small $^{21}$ University of Chinese Academy of Sciences, Beijing, 100049, China\vspace*{-2pt} \\ 
}

\date{Accepted XXX. Received YYY; in original form ZZZ}

\pubyear{2020}

\begin{document}
\label{firstpage}
\pagerange{\pageref{firstpage}--\pageref{lastpage}}
\maketitle

\begin{abstract}
We present the anisotropic clustering of emission line galaxies (ELGs) from the Sloan Digital Sky Survey IV (SDSS-IV) extended Baryon Oscillation Spectroscopic Survey (eBOSS) Data Release 16 (DR16). Our sample is composed of 173,736 ELGs covering an area of 1170 deg$^2$ over the redshift range $0.6 \leq z \leq 1.1$. We use the Convolution Lagrangian Perturbation Theory in addition to the Gaussian Streaming Redshift-Space Distortions to model the Legendre multipoles of the anisotropic correlation function. We show that the eBOSS ELG correlation function measurement is affected by the contribution of a radial integral constraint that needs to be modelled to avoid biased results. To mitigate the effect from unknown angular systematics, we adopt a modified correlation function estimator that cancels out the angular modes from the clustering. At the effective redshift, $z_{\rm eff}=0.85$, including statistical and systematical uncertainties, we measure the linear growth rate of structure $f\sigma_8(z_{\rm eff}) = 0.35\pm0.10$, the Hubble distance $D_H(z_{\rm eff})/r_{\rm drag} = 19.1^{+1.9}_{-2.1}$ and the comoving angular diameter distance $D_M(z_{\rm eff})/r_{\rm drag} = 19.9\pm1.0$. These results are in agreement with the Fourier space analysis, leading to consensus values of: $f\sigma_8(z_{\rm eff}) = 0.315\pm0.095$, $D_H(z_{\rm eff})/r_{\rm drag} = 19.6^{+2.2}_{-2.1}$ and $D_M(z_{\rm eff})/r_{\rm drag} = 19.5\pm1.0$, consistent with $\Lambda$CDM model predictions with Planck parameters.
\end{abstract}




\section{Introduction}  \label{sec:intro}
For the last 20 years, physicists have known that the expansion of the Universe is accelerating \citep{Riess:1998aa, Perlmutter:1999aa}, but not why this is happening, although the mechanism has been given a name: dark energy. In the simplest mathematical model, the acceleration is driven by a cosmological constant $\Lambda$, inside Einstein's field equations of General Relativity (GR), and this model is referred to as the standard model of cosmology or the $\Lambda$CDM model. Precise measurements of the Cosmic Microwave Background \citep{Planck-Collaboration:2016aa}, combined with the imprint of the Baryon Acoustic Oscillations (BAO) in the clustering of galaxies \citep{Eisenstein:2005aa,Cole:2005aa}, in particular for those from the Baryon Oscillation Spectroscopic Survey (BOSS), \citep{Alam:2017aa} indicate that dark energy contributes 69\% of the total content of the Universe, while dark and baryonic matter only contribute 26\% and 5\% respectively.

Measurements of BAO are only one component of the information available from a galaxy survey. The observed large-scale distribution of galaxies depends on the distribution of matter (which includes the BAO signal), the link between galaxies and the mass known as the bias, geometrical effects that project galaxy positions into observed redshifts and angles, and Redshift-Space Distortions (RSD).

RSD arise because the measured redshift of a galaxy is affected by its own peculiar velocity, a component that arises from the growth of cosmological structure. These peculiar velocities lead to an anisotropic clustering, as first described in the linear regime by \cite{Kaiser:1987aa}. In linear theory, the growth rate of structure $f$ is often parameterised using:
\begin{equation}\label{eq:fz}
f(a)=\cfrac{d \ln D(a)}{d \ln a}
\end{equation}
where $D(a)$ is the linear growth function of density perturbations and $a$ is the scale factor. In practice, RSD provide measurements of the growth rate via the quantity $f(z)\sigma_8(z)$, where $\sigma_8(z)$ is the amplitude of the matter power spectrum at 8\mpc \citep{Song:2009aa}. In the framework of General Relativity, the growth rate $f$ is related to the total matter content of the Universe $\Omega_m$ through the generalized approximation \citep{Peebles:1980aa}:
\begin{equation}
f(z) \simeq \Omega_m(z)^\gamma
\end{equation}
where the exponent $\gamma$ depends on the considered theory of gravity and is predicted to be $\gamma \simeq 0.55$ in GR \citep{Linder:2007aa}. Therefore by measuring the growth rate of structure in the distribution of galaxies as function of redshift, we can put constrains on gravity, and test if dark energy could be due to deviations from GR \citep{Guzzo:2008aa}. 

BAO and RSD measurements are highly complementary, as they allow both geometrical and dynamical cosmological constraints from the same observations. In addition, BAO measurements break a critical degeneracy affecting RSD measurements: clustering anisotropy arises both due to RSD and also if one assumes a wrong cosmology to transform redshifts to comoving distances. The latter is known as the Alcock-Paczynski (AP) effect \citep{Alcock:1979aa} and generates distortions both in the angular and radial components of the clustering signal. The AP effect shifts the BAO peak, while leaving the RSD signal unaffected, and hence anisotropic BAO measurements break the AP-RSD degeneracy and enhance RSD measurements.

Using BAO and RSD measurements, spectroscopic surveys of galaxies are now amongst the most powerful tools to test our cosmological models and in particular to probe the nature of dark energy. Up until now, the most powerful survey has been BOSS~\citep{Dawson:2013aa}, 
which made two $\sim$1\% precision measurements of the BAO position at $z=0.32$, and $z=0.57$ \citep{Alam:2017aa}, coupled with two $\sim8$\% precision measurements of $f\sigma_8$ from the RSD signal. The extended Baryon Oscillation Spectroscopic Survey (eBOSS; \citealt{Dawson:2016aa}) program is the follow-up for BOSS in the fourth generation of the Sloan Digital Sky Survey (SDSS; \citealt{Blanton:2017aa}). With respect to BOSS, it explores large-scale structure at higher redshifts, covering the range $0.6<z<2.2$ using four main tracers: Luminous Red Galaxies (LRGs), Emission Line Galaxies (ELGs), quasars used as direct tracers of the density field, and quasars from whose spectra we can measure the Ly$\alpha$ forest. In this paper we present RSD measurements obtained from ELGs in the final sample of eBOSS observations: Data Release 16 (DR16). Using the first two years of data released as DR14 \citep{Abolfathi:2018aa}, BAO and RSD measurements have been made using the LRGs \citep{Bautista:2018aa,Icaza-Lizaola:2019aa} and quasars \citep{Ata:2018aa,Gil-Marin:2018aa,Zarrouk:2018aa}, but not the ELG sample, which was not complete for that data release.

The eBOSS ELG sample, covering $0.6<z<1.1$, is fully described in \citet{raichoor20a}. As well as allowing high redshift measurements, this sample is important because it is a pathfinder sample for future experiments as DESI \citep{DESI-Collaboration:2016aa,DESI-Collaboration:2016ab}, \textit{Euclid} \citep{Euclid:2011aa}, PFS \citep{Sugai:2012aa,Takada:2014aa}, or \textit{WFIRST} \citep{Dore:2018aa} which will also focus on ELGs. We analyse the first three even Legendre multipoles of the anisotropic correlation function to measure RSD and present a RSD+BAO joined measurement. A companion paper describes the BAO \& RSD measurements made in Fourier-space \citep{demattia20a}, while BAO measurements in configuration space are included in \citet{raichoor20a}. A critical component for interpreting our measurements is the analysis of fast mocks catalogues \citep{lin20a,zhao20a}. We also use mocks based on N-body simulations to understand the systematic errors \citep{Alam2020,Avila2020}.

The eBOSS ELG sample suffers from significant angular fluctuations because it was selected from imaging data with anisotropic properties, which imprint angular patterns \citep{raichoor20a} such that we cannot reliably use angular modes to measure cosmological clustering. Traditionally, when the modes affected are known they are removed from the measurement either by assigning weights to correct for observed fluctuations \citep{Ross:2011aa}, or by nullifying those modes \citep{Rybicki:1992aa}. In fact, these approaches are mathematically equivalent \citep{Kalus:2016aa}. In the extreme case that we do not know the contaminant modes, one can consider nulling all angular modes. This can be achieved by matching the angular distributions of the galaxies and mask - an extreme form of weighting \citep{Burden:2017aa,Pinol:2017aa} or, in the procedure we adopt, by using a modified statistic designed to be insensitive to angular modes. 

The ELG studies described above are part of a coordinated release of the final eBOSS measurements of BAO and RSD in all samples including the LRGs over $0.6<z<1.0$ \citep{LRG_corr,gil-marin20a} and quasars over $0.8<z<2.2$ \citep{hou20a,neveux20a}. For these samples, the construction of data catalogs is presented in \citet{ross20a,lyke20a}, and N-body simulations for assessing systematic errors \citep{rossi20a,smith20}. At the highest redshifts ($z>2.1$), our release includes measurements of BAO in the Lyman-$\alpha$ forest \citep{2020duMasdesBourbouxH}. The cosmological interpretation of all of our results together with those from other cosmological experiments is found in \citet{eBOSS_Cosmology}. A SDSS BAO and RSD summary of all tracers measurements and their full cosmological interpretation can be found on the SDSS website\footnote{\url{https://www.sdss.org/science/final-bao-and-rsd-measurements/}.\\
\url{https://www.sdss.org/science/cosmology-results-from-eboss/}}.

We summarise the ELG data used in Section~\ref{sec:data}, and the mock catalogues in Section~\ref{sec:mocks}. The analysis method that nulls angular modes, designed to reduce systematic errors is described in Section~\ref{sec:method}. The model fitted to the data is presented in Section~\ref{sec:model}. Section~\ref{sec:analysis} validates with the mock catalogues our chosen modelling and the analysis method to reduce angular contamination. Finally, we present our results in Section~\ref{sec:results}, and conclusions in Section~\ref{sec:conclusions}.

\section{Data}  \label{sec:data}
In this Section, we summarise the eBOSS ELG large-scale structure catalogues which are studied in this paper and refer the reader to \citet{raichoor20a} for a complete description. 
The eBOSS ELG sample was selected on the $grz$-bands photometry of intermediate releases (DR3, DR5) of the DECam Legacy Survey imaging (DECaLS), a component of the DESI Imaging Legacy Surveys \citep{Dey:2019aa}.
This photometry is more than one magnitude deeper than the SDSS photometry.
The target selection is slightly different in the two caps, as the DECaLS photometry is deeper in the SGC than in the NGC.
The selected targets were then spectroscopically observed during approximately one hour with the BOSS spectrograph \citep{Smee:2013aa} at the 2.5-meter aperture Sloan Foundation Telescope at Apache Point Observatory in New Mexico \citep{Gunn:2006aa}.
We refer the reader to \cite{Raichoor:2017aa} for a detailed description of the target selection and spectroscopic observations.

The catalogues used contain 173,736 ELGs with a reliable spectroscopic redshift, $z_{\rm spec}$, between 0.6 and 1.1, within a footprint split in two caps, the North Galactic Cap (NGC) and South Galactic Cap (SGC).
For the spectroscopic observations, each cap is split into two 'chunks', which are approximately rectangular regions where the tiling completeness is optimized.
Table \ref{ELG_stats} presents the number of used $z_{\rm spec}$ and the effective area, i.e. the unmasked area weighted by tiling completeness, for each cap and for the combined sample; it also reports redshift information if one restricts to $0.7<z_{\rm spec}<1.1$, 
as this range is used in the RSD analysis (see Section \ref{sec:results}).

\begin{table}
	\centering
	\begin{tabular}{lccccc}
		\hline 
		& $z_{\rm min}$ & $z_{\rm max}$& NGC & SGC & ALL\\
		\hline 
		Effective area [deg$^2$] & - & - & 369.5 & 357.5 & 727.0 \\
		Reliable redshifts & 0.6 & 1.1 & 83,769 & 89,967 & 173,736 \\
		& 0.7 & 1.1 & 79,106 & 84,542 & 163,648 \\
		Effective redshift & 0.6 & 1.1 & 0.849 & 0.841 & 0.845\\
		& 0.7 & 1.1 & 0.860 & 0.853 & 0.857\\
	    \hline
	\end{tabular} 
	\caption{Effective area and number of reliable redshifts per Galactic cap and in the combined ELG sample.}
	\label{ELG_stats}
\end{table}

\begin{figure}
\centering
\includegraphics[width=0.95\columnwidth]{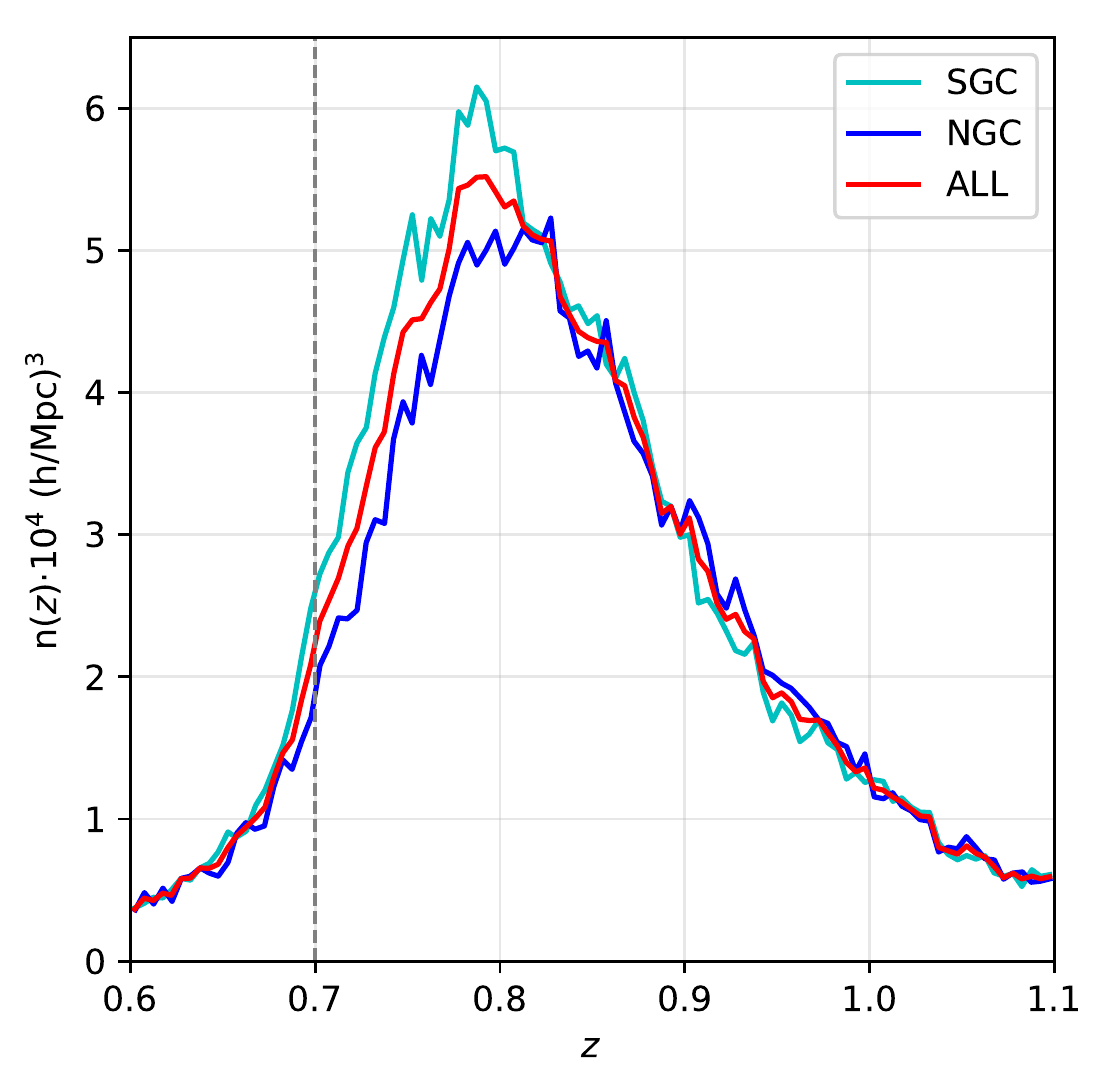}
\caption{\label{dndchi} 
Redshift density of the eBOSS ELG sample per Galactic cap and for the combined sample.}
\end{figure}

Different weights and angular veto masks are applied to data, to correct for variations of the survey selection function, as described in more details in~\citet{raichoor20a}.
In particular, weights are introduced to correct for fluctuations of the ELG density with imaging quality (systematic weight $w_{\rm sys}$), to account for fibre collisions (close-pair weight $w_{\rm cp}$) and to  correct for redshift failures ($w_{\rm noz}$ weight).
Figure \ref{dndchi} shows the 
redshift density ($n(z)$) of the ELG sample for the two Galactic caps and the combined sample.
The more numerous $z_{\rm spec}<0.8$ ELGs in the SGC is a consequence of the target selection choice to explore a larger box in the $g-r$ vs. $r-z$ colour-colour diagram, enabled by the deeper photometry there \citep{Raichoor:2017aa}.
As in previous BOSS/eBOSS analyses \citep[e.g.][]{Anderson:2014aa}, we also define inverse-variance $w_{\rm FKP}$ weights, $w_{\rm FKP} = 1/(1+n(z) \cdot P_0)$ \citep{Feldman:1994aa}, with $P_0 = 4000$ $h^{-3}$ Mpc$^3$.

Consistently with the other eBOSS analyses, we define the effective redshift ($z_{\rm eff}$) of the ELG sample as the weighted mean spectrscopic redshift of galaxy pairs ($z_i,z_j$):
\begin{equation}
z_{\rm eff} = \frac{\sum_{i,j} w_{\rm{tot},i} w_{\rm{tot},j} (z_i + z_j)/2}{\sum_{i,j} w_{\rm{tot},i} w_{\rm{tot},j}},
\label{eq:zeff}
\end{equation}
where $w_{\rm tot} = w_{\rm sys} \cdot w_{\rm cp} \cdot w_{\rm noz}\cdot w_{\rm FKP}$ and the sum is performed over all galaxy pairs between 25 $h^{-1}$ Mpc and 120 $h^{-1}$ Mpc.
We report in Table \ref{ELG_stats} the different $z_{\rm eff}$ values for the NGC, SGC, and combined sample for $0.6<z_{\rm spec}<1.1$ and $0.7<z_{\rm spec}<1.1$.

A random catalogue of approximately 40 times the data density is created to account for the survey selection function of the weighted data. Angular coordinates of random objects are uniformly distributed and those objects outside the footprint and masks are rejected. Random objects are assigned data redshifts, according to the shuffled scheme introduced in~\citet{Ross:2012aa}. As described in~\citet{raichoor20a}, this was done per chunk, in separate sub-regions of approximately constant imaging depth, in order to account for the fact that targets selected in regions of shallower imaging have lower redshifts on average.

As shown in~\citet{de-Mattia:2019aa} and \citet{demattia20a}, using the shuffled-z scheme leads to the suppression of radial modes and impacts the multipoles of the measured correlation function. This effect has to be modelled, a point we develop in Section~\ref{RICsection}.

Despite the different corrections, the eBOSS ELG sample still suffers from significant angular systematics (see Section~\ref{2pcf_sec}), likely due to unidentified systematics in the imaging data used to select ELG targets, a point further discussed in~\citet{demattia20a}. This triggered our using of the modified correlation function described in Section~\ref{sec:modif2PCF} to cancel the angular modes.

\section{Mocks} \label{sec:mocks}
In this Section, we briefly describe the mock catalogues used in the analysis. Those mock catalogues are of two types: approximate mocks to estimate the covariance matrix and validate the pipeline analysis and precise N-body mocks to validate the model.

\subsection{EZmocks}\label{sec:EZmocks}
A thousand EZmock catalogues for each Galactic cap are used to estimate the covariance matrices for parameter inference. These mocks rely on the Zel'dovich approximation \citep{Zeldovich:1970aa} to generate the dark matter density field, with $1024^3$ grids in a $5^3\,h^{-3}\,{\rm Gpc}^3$ comoving box.
ELGs are then populated using an effective galaxy bias model, which is directly calibrated to the 2- and 3-point clustering measurements of the eBOSS DR16 ELG sample \citep{Chuang:2015aa, zhao20a}.
The cosmology used to generate the EZmocks is a flat $\Lambda$CDM model with:
\begin{equation}
\begin{split}
\quad\quad h=0.6777,\, \,  \Omega_{\rm m}&=0.307115,\, \, \Omega_{\rm b}=0.048206,&\\ \, \, \sigma_8&=0.8225,\, \, n_{\rm s}= 0.9611&
\end{split}
\label{eq:cosmoEZ}
\end{equation}

To account for the redshift evolution of ELG clustering, the EZmock simulations are generated with seven redshift snapshots. These snapshots are converted to redshift space, to construct slices with the redshift ranges of $(0.6, 0.7)$, $(0.7, 0.75)$, $(0.75, 0.8)$, $(0.8, 0.85)$, $(0.85, 0.9)$, $(0.9, 1.0)$, and $(1.0, 1.1)$.
The slices are then combined, and the survey footprint and veto masks are applied to construct light-cone mocks that reproduce the geometry of the data. 

Depending on how the radial and angular distributions of the eBOSS data are migrated to the light-cone mocks, two sets of EZmocks -- without systematics and with systematics -- are generated. For the mocks \textit{without systematics}, only the radial selection is applied, to mimic the redshift evolution of the eBOSS ELG number density. Moreover, the radial selections are applied separately for different chunks, since their spectroscopic properties are different \citep{raichoor20a}. 
Thus, the only observational effect applied on the angular distribution of the EZmocks without systematics is the footprint geometry and veto masks.

The EZmocks \textit{with systematics}, however, encode observational systematic effects, namely angular photometric systematics, fibre collisions, and redshift failures. For example, a smoothed angular map of galaxy positions is extracted directly from the data, and applied to the mocks. The photometric and spectroscopic effects are then corrected by the exact same weighting procedure as in data~\citep[see][for details]{demattia20a, zhao20a}. In particular, mock data redshifts are randomly assigned to mock random catalogues with the 'shuffled-z' scheme in chunks of homogeneous imaging depth (using the depth map of the eBOSS data).
Moreover, a smoothed angular map of galaxy positions is extracted directly from the data, and applied to the mocks. The photometric and spectroscopic effects are then corrected by the exact same weighting procedure as in data~\citep[see][for details]{demattia20a, zhao20a}.

In this study, we further use two variants of the EZmocks with systematics, which differ in their random catalogues. 
The redshift distribution of the random objects should reflect the radial survey selection function of the corresponding galaxy catalogue. This can be achieved in two ways, either by sampling the random redshifts based on the true radial selection function $n(z)$ of data, or by taking directly the shuffled redshifts from the galaxy catalogue. We dub these two schemes `sampled-z' and `shuffled-z', respectively. For the  EZmocks with systematics only the `shuffled-z' randoms are used.

\subsection{N-body mocks}\label{nbody}
The eBOSS ELG sample significantly differs from the other eBOSS tracers from a galaxy formation point-of-view. These galaxies are sites of active star formation with various astrophysical processes at play, such as the consumption of gas or the effect of the local environment.
This means the kinematical properties of eBOSS ELGs could be different from those of the underlying dark matter haloes. One must thus test the robustness of any cosmological inference against galaxy formation physics.
To do so, we tested our model against a wide variety of eBOSS ELG mock catalogues which include accurate non-linear evolution of dark matter and various deviations in galaxy kinematics from the underlying dark matter distribution. These tests are described in detailed in a companion paper \citet{Alam2020}. Briefly, we employ two different N-body simulations, the {\sc Multi Dark Planck} \citep[MDPL2;][]{Klypin:2016aa} and the {\sc Outer Rim} 
\citep[OR;][]{Heitmann:2019aa}. \\ \\
The MDPL2 simulation provides a halo catalogue produced with the Rockstar halo finder \citep{Behroozi:2013aa} in a cubic box of 1 $h^{-1}$Gpc using a flat $\Lambda$CDM cosmology with parameters: 
\begin{equation}
\begin{split}
\quad\quad h=0.6777,\, \,  \Omega_{\rm m}&=0.307115,\, \, \Omega_{\rm b}=0.048206,&\\ \, \, \sigma_8&=0.8228,\, \, n_{\rm s}= 0.9611&
\end{split}
\label{eq:cosmoMD}
\end{equation}
\\
The OR simulation provides a halo catalogue produced with the Friends of Friends halo finder of \cite{Davis:1984aa} in a cubic box of 3 $h^{-1}$Gpc using a flat $\Lambda$CDM cosmology with parameters: 
\begin{equation}
\begin{split}
\quad\quad h=0.71,\, \,  \Omega_{\rm CDM}h^2&=0.1109,\, \, \Omega_{\rm b}h^2=0.02258,&\\ \, \, \sigma_8&=0.8,\, \, n_{\rm s}= 0.963&
\end{split}
\label{eq:cosmoOU}
\end{equation}
\\ 
Three different parametrisations for the shape of the mean HOD (Halo Occupation Distribution) of central galaxies are used. The first parametrisation called SHOD is the standard HOD model where at least one central galaxy of a given type is found in massive enough dark matter haloes. Although this model is more appropriate for modelling magnitude or stellar mass selected samples \citep{Zheng:2005aa,White:2011aa}, it can be modified to account for the incompletness in mass of a sample such as the ELG one. The second parametrisation is called HMQ which essentially quenches galaxies at the centre of massive haloes and suppresses the presence of ELGs in the center of haloes, as suggested by observations and models of galaxy formation, and hence should provide more realistic realisation of star-forming ELGs \citep{Alam:2019aa}. The third parametrisation, 
called SFHOD, 
accounts for the incompletness of the ELG sample by modelling central galaxies with an asymmetric Gaussian \citep{Avila2020}. Such a shape is based on the results from the galaxy formation and evolution model presented in \citet{GonzalezPerez:2018aa}. In each of these models, besides the shape of the mean HOD, other aspects have been varied to mimic different possible baryonic effects over the ELGs distribution such as the satellite distribution, infalling velocities, the off-centring of central galaxies and the existence of assembly bias.

In total 22 MDPL2 mocks were available, with 11 types of mocks for each of the SHOD and HMQ models.
OR mocks encompassed 6 out of the 11 same types for each model, and five SFHOD models with assumptions that enhance the parameter space explored by the SHOD and HMQ ones are selected.

As the MDPL2 cosmology is close to our fiducial BOSS cosmology (Equation \ref{eq:cosmoBOSS}), we use the latter to analyse the MDPL2 mocks.
We analyse the OR mocks with their own cosmology (Equation \ref{eq:cosmoOU}). For the covariance matrix, we use an analytical covariance as defined in \cite{Grieb:2016aa}.

\section{Method}  \label{sec:method}
\subsection{The two-point correlation function} \label{2pcf_sec}

To compute galaxy pair separations of data and EZmocks, 
observed redshifts need first to be converted into comoving distances. To do so, we use the same flat $\Lambda$CDM fiducial cosmology as in BOSS DR12 analysis~\citep{Alam:2017aa}: 
\begin{equation}
\begin{split}
\quad\quad h = 0.676, \,\, 
\Omega_{\rm m} = 0.31,  \,\, 
\Omega_\Lambda = 0.69, \,\, \Omega_{\rm b}h^2 = 0.022, \\	
\sigma_8 = 0.8, \,\,	
n_{\rm s} = 0.97,	\,\, 
\sum m_{\nu} = 0.06\ \rm{eV} \qquad
\end{split}
\label{eq:cosmoBOSS}
\end{equation}

Afterwards, in order to quantify the anisotropic galaxy clustering in configuration space, one usually resorts to the two-point correlation function $\xi$ (2PCF), which is defined as the excess probability of finding a pair of galaxies separated by a certain vector distance $\vect{s}$ with respect to a random uniform distribution. 
In the next Sections, we refer to that 2PCF as the 'standard 2PCF'.

An unbiased estimate $\hat{\xi}$ of the correlation function $\xi$ can be computed for a line of sight separation $s_\parallel$ and transverse separation $s_\perp$, using the \citet[][LS]{Landy:1993aa} estimator: 
\begin{equation}
\hat{\xi}(\sperp,\spara)  = \frac{DD(\sperp,\spara)-2DR(\sperp,\spara)+RR(\sperp,\spara)}{RR(\sperp,\spara)},
\label{eq:LS}
\end{equation}
where $DD$, $DR$, and $RR$ are the normalised galaxy-galaxy, galaxy-random, and random-random pair counts, respectively.
The pair separation can also be written in terms of $s$ and $\mu=s_\parallel/s = \cos(\theta)$, where $\theta$ is the angle between the pair separation vector $\vect{s}$ and the line of sight.

Projecting on the basis of Legendre polynomials, the two-dimensional correlation function is compressed into multipole moments of 
order $l$ \citep{Hamilton:1992aa}:
\begin{equation} \label{multipole_def}
\begin{split} 
\xi_{\ell}(s) &\equiv \cfrac{2\ell+1}{2}\displaystyle\int_{-1}^1 d\mu\xi(s, \mu)P_l(\mu)\\
		 &= \cfrac{2\ell+1}{2}\displaystyle\int_{0}^\pi d\theta\sqrt{1-\mu^2}\xi(s_\perp,s_\parallel)P_{\ell}(\mu)
\end{split} 
\end{equation}
where $P_{\ell}(\mu)$ is the Legendre polynomial of order $\ell$.

Equations \ref{multipole_def} 
are integrated over a spherical shell of radius $s$, while measurements of $\hat{\xi}(s_\perp,s_\parallel)$  are performed in bins of width $\Delta s$ in $s_\perp,s_\parallel$. 
Converting the last integral in Equation \ref{multipole_def} to sums over bins leads to the following definition of the 
estimated multipoles of the correlation function \citep{Chuang:2013aa}:\\
\begin{equation}\label{eq:effmult}
\hat{\xi}_{\ell}(s) \equiv \cfrac{(2\ell+1)}{2} \cfrac{\pi}{n}
\displaystyle \sum_{i=1}^{n} 
\sqrt{1-\mu_i^2}\hat{\xi}(s^i_\perp,s^i_\parallel)P_{\ell}(\mu_i)
\end{equation}
where the sum extends over $n$ bins in $s_\perp,s_\parallel$ obeying:
\begin{equation*}
s-\frac{\Delta s}{2} < \sqrt{s_\parallel^2 + s_\perp^2} < s+\frac{\Delta s}{2}  
\end{equation*}

We use the public code \textsc{cute} \citep{Alonso:2012aa} to evaluate the LS estimator of the correlation function from the data and \textsc{FCFC} code \citep[fast correlation function calculator;][]{Zhao:2020ab} for the mocks: both codes provide consistent measurements. For both mocks and data, we then compute the first even multipoles, $\hat{\xi_0}, \hat{\xi_2}$ and $\hat{\xi_4}$, in bins of width $\Delta s=8$\mpc for each cap separately.
The combined multipoles over both caps, referred as ALL, are computed by averaging the NGC and SGC multipoles, weigthed by their respective effective areas, $A_{\rm NGC}$ and $A_{\rm SGC}$:
\begin{equation}
\hat{\xi}^{\rm ALL}_{\ell}(s_\perp,s_\parallel) = \cfrac{\hat{\xi}^{\rm NGC}_{\ell}(s_\perp,s_\parallel)A_{\rm NGC} + \hat{\xi}^{\rm SGC}_{\ell}(s_\perp,s_\parallel)A_{\rm SGC}}{A_{\rm NGC} + A_{\rm SGC}}
\end{equation}
\begin{figure*}
\centering
\includegraphics[width=1.99\columnwidth]{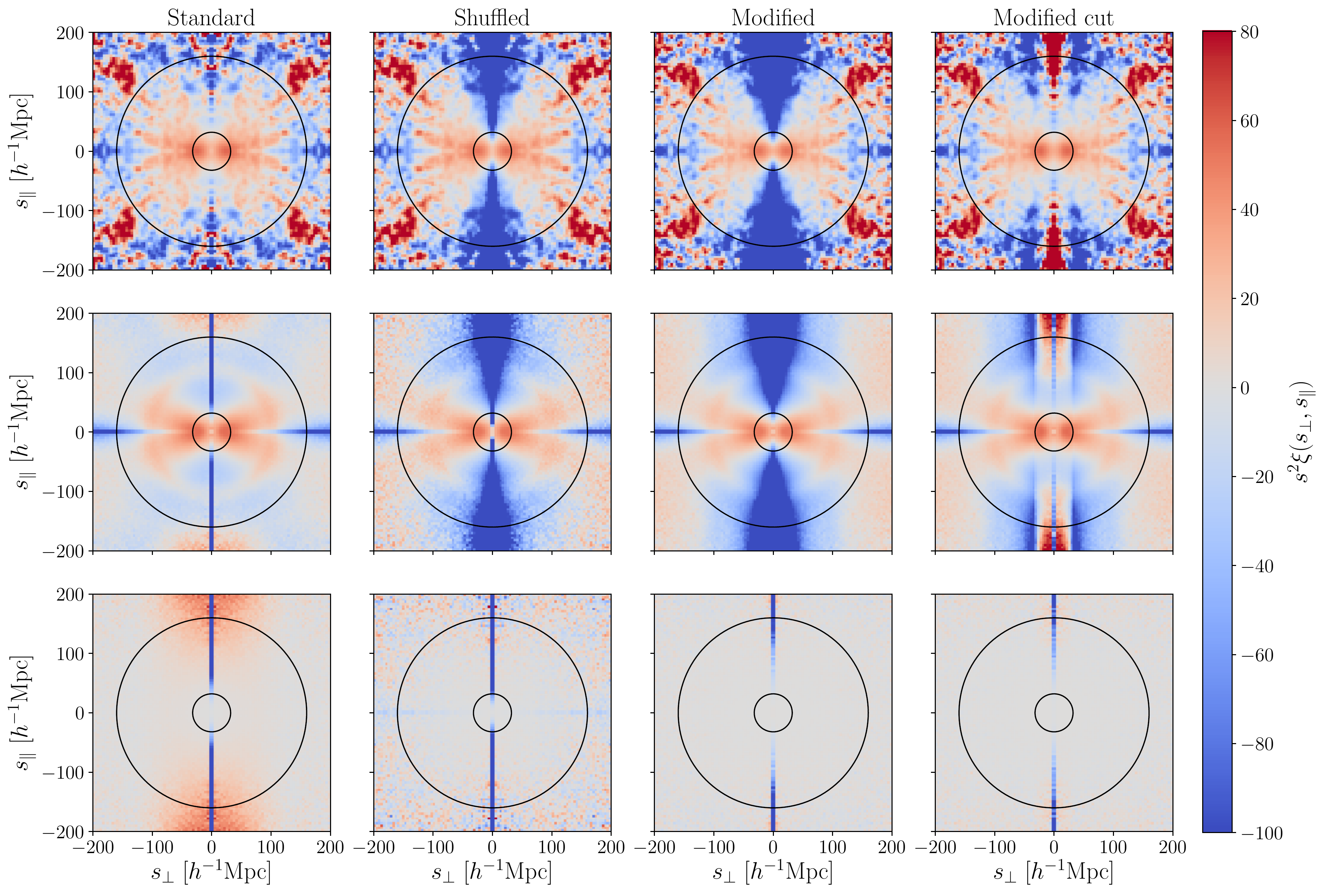}
\caption{\label{fig:2DdataMocks} Two-dimensional correlation function in the directions perpendicular and parallel to the line-of-sight. Panels from left to right are for: standard 2PCF (Equation \ref{eq:LS}), shuffled 2PCF (Equation \ref{eq:shuffle2PCF}), modified 2PCF with no cut (Equation \ref{eq:mod2pcf}) and modified 2PCF with a cut (Equation \ref{eq:mod2pcffinal}). The top row displays the measurement from the eBOSS ELG data sample, the middle row displays the mean of the 1000 'shuffled-z' EZmocks with systematics (100 mocks for the shuffled 2PCF), and the bottom row shows the difference between the mean of the 1000 'shuffled-z' EZmocks with and without systematics (100 mocks for the shuffled 2PCF).
For the modified 2PCF, all parameters are taken at their fiducial values (see text).
The black circles illustrate our fiducial fitting range in $s$ for the multipoles.
}
\end{figure*}

The top and middle left panels of Figure~\ref{fig:2DdataMocks} show the standard 2PCF of the data and 
mean of the 1000 'shuffled-z' EZmocks with systematics, respectively. The squashing effect due to RSD can be observed for both data and EZmocks; the BAO signal is clearly visible in the EZmocks, but not in data, because of the overall low statistics, as seen in \citet{raichoor20a}. For the mocks, and for data to a lesser extent, we see a negative clustering at $\spara \sim 0$:
this is due to the 'shuffled-z' scheme adopted to assign redshifts to random objects, which creates an excess of $DR$ and $RR$ pairs at those values. The bottom left panel of Figure~\ref{fig:2DdataMocks} displays the difference between the mean of the 1000 'shuffled-z' EZmocks without and with systematics: the systematics show up mostly at small $\sperp$ (radial, due to spectroscopic observations) and large $\spara$ (angular, due to the imaging systematics).

\begin{figure}
\includegraphics[width=0.95\columnwidth]{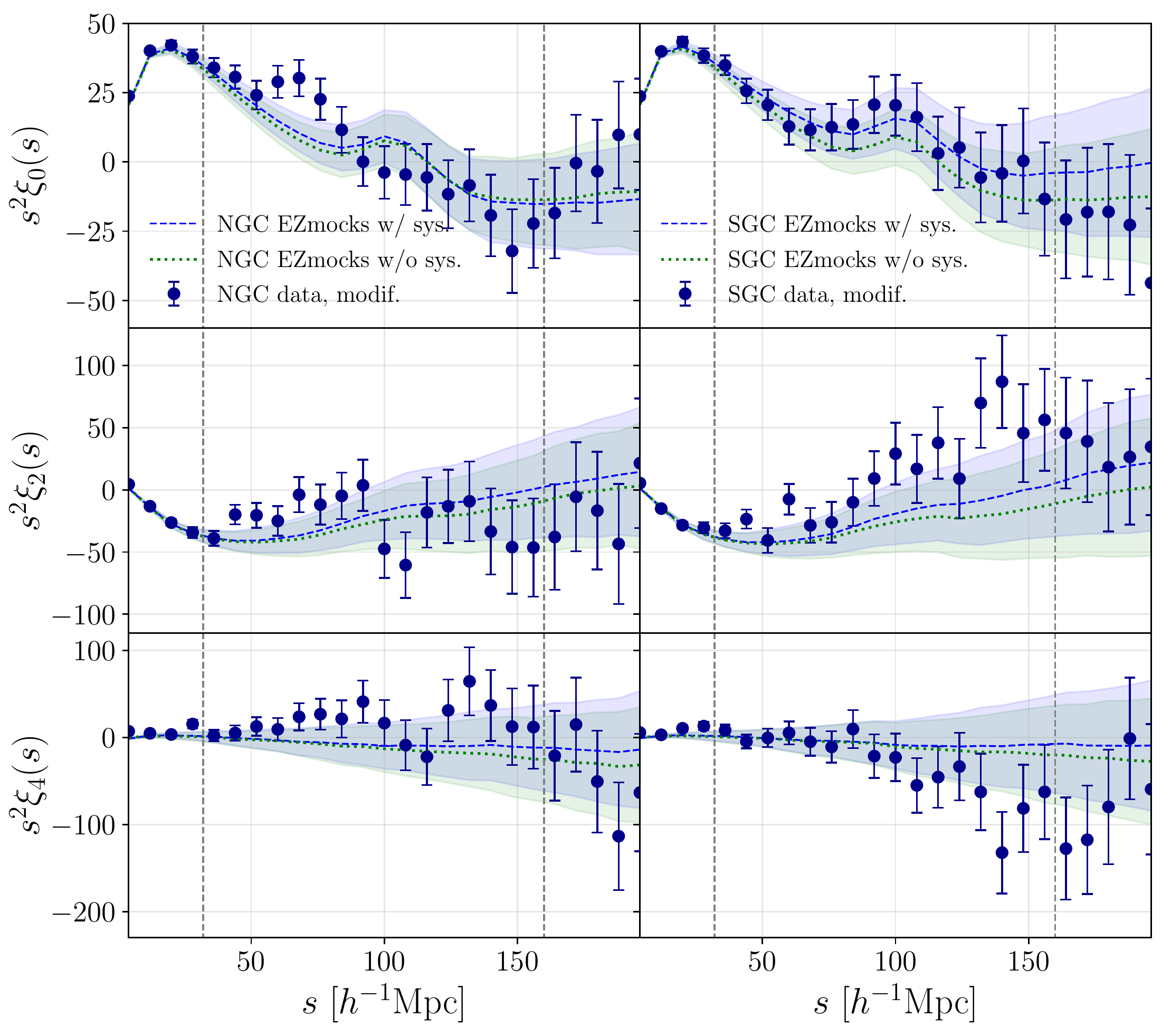}
\caption{\label{fig:datacorr} Multipoles of the standard 2PCF as measured from the eBOSS ELG data sample in each cap and from the mean of the shuffled-z EZmocks with and without systematics. The bands represent the one sigma dispersion around the mean of the mocks. Errors on data points come from one sigma dispersion of mocks with systematics. Vertical 
dashed lines 
define the baseline fitting 
range.
}
\end{figure}

Figure~\ref{fig:datacorr} shows the standard 2PCF multipoles for the data and for the 'shuffled-z' EZmocks with or without systematics, separately for the NGC and the SGC. Adding systematics to the EZmocks improves the agreement with data, especially for the monopole in the SGC and for the quadrupole in both caps. The overall agreement is satisfactory. However, there are remaining discrepancies between the data and the EZmocks with systematics, the most significant ones being at intermediate scales, $\sim40-80$\mpc, in the NGC for the monopole and the quadrupole. As detailed in the next Section, those are likely due to remaining angular systematics in the data.

\subsection{Modified 2PCF} \label{sec:modif2PCF}

In order to mitigate those systematics 
in our RSD analysis, we use a modified 2PCF built on the standard $\xi(\sperp,\spara)$ for the model and $\hat{\xi}(\sperp,\spara)$ for data and mocks. 
Actually, as will be shown in Section \ref{sec:mitigatesys} with the EZmocks, fitting the standard 2PCF multipoles $\hat{\xi}_{\ell}(s)$ does not allow us to to recover unbiased cosmological parameters when data-like systematics are included in the mocks -- and corrected as in data. The principle of the modified 2PCF is thus to null the angular modes from the clustering.

Our approach builds on the method presented in \citet{Burden:2017aa} designed for the DESI survey, in which they proposed a modification of the correlation function that nulls the angular modes from the clustering. \citet{Burden:2017aa} introduce the shuffled 2PCF which is a modification of the LS estimator from Equation~\ref{eq:LS}: 
\begin{equation}
\hat{\xi}^{\rm shuff}(\sperp,\spara) = \frac{DD(\sperp,\spara) -2DS(\sperp,\spara)+SS(\sperp,\spara)}{RR(\sperp,\spara)},
\label{eq:shuffle2PCF}
\end{equation}
where $S$ stands for a random catalog built with random picks of the data angular positions and with a radial distribution following the data one (according to the 'shuffled-z' scheme in our case).
Using such a random $S$ catalog, with the same angular clustering as that in the galaxy catalog, implies that angular modes are removed in the shuffled 2PCF, at the cost of an overall loss of information. 
Second column of Figure~\ref{fig:2DdataMocks} shows the two-dimensional shuffled 2PCF of data (first row) and the mean (second row) of 100 EZmocks with systematics: angular signal at small $\sperp$ and large $\spara$ are removed. On the bottom row of the second column of Figure~\ref{fig:2DdataMocks}, we present the difference between the mean of EZmocks with and without systematics. As most systematics are removed compared to the standard 2PCF, this suggests that the nature of the uncorrected systematics mostly comes from angular signal and that the shuffled 2PCF removes them.

A model for the shuffled 2PCF was also presented in \citet{Burden:2017aa} and shown to provide an unbiased isotropic BAO measurement.
However, a more advanced modelling is required for a RSD analysis, as we are measuring anisotropic information from the monopole, quadrupole and hexadecapole.
The model of \citet{Burden:2017aa} involves subtracting terms integrated over the line of sight which thus include scales for which the RSD model may be invalid (see Section \ref{sec:model}). Such small scales will be discarded from our fits. For that reason, we do not use the shuffled 2PCF for our measurements on data and mocks, but rely on a modified 2PCF where we can control the boundaries of integration for both data and model.
The modified 2PCF we adopt is based on:
\begin{subequations}
\begin{align}
\xi^{\rm mod}(\sperp,\spara) = & 
\; \xi(\sperp,\spara) \\
& - 2 \int_{-\spara^{\rm max}}^{\spara^{\rm max}}\xi(\sperp,\spara') \bar{n}(\chi_{\rm mod}+\spara'/2)d\spara' \label{eq:mod2pcf_2}\\
& + \int_{0}^{\infty} \bar{n}^2(\chi)d\chi 
    \int_{-\spara^{\rm max}}^{\spara^{\rm max}} \xi(\sperp,\spara')d\spara', \label{eq:mod2pcf_3}
\end{align}
\label{eq:mod2pcf}
\end{subequations}
where $\bar{n}(\chi)$ is the normalized data radial density as a function of the comoving line-of-sight distance $\chi$ and $\chi_{\rm mod}$ is the comoving line-of-sight distance at a given redshift $z_{\rm mod}$, defined hereafter. $\spara^{\rm max}$ is the maximum parallel scale included in the correction.
Equation~\ref{eq:mod2pcf_2} corresponds to the cross-correlation between the three-dimensional overdensity and the projected angular overdensity and Equation~\ref{eq:mod2pcf_3} corresponds to the angular correlation function. We provide more 
details about  Equation~\ref{eq:mod2pcf} in Appendix~\ref{sec:appendix_modifcorr}.
The third column 
of Figure~\ref{fig:2DdataMocks} illustrates the modified 2PCF defined in Equation~\ref{eq:mod2pcf}: it clearly shows its efficiency to remove the angular clustering in the data (top row) and in the mocks (middle row), with as a consequence a significant removal of the angular systematics. This can also be seen on the third bottom panel, where the systematics included are almost completely cancelled. We note that the modified 2PCF and the shuffled 2PCF are very similar.

In our implementation, we use $\spara^{\rm max}=190$\mpc and $z_{\rm mod}=0.83$ as baseline parameters. Both quantities are treated as parameters and chosen to minimise the systematics.
One can note in Equations~\ref{eq:mod2pcf_2} and \ref{eq:mod2pcf_3} that the integration does not depend on the value of $s = \sqrt{\sperp^2+\spara'^2}$.
However, since the CLPT-GS model is not valid on small scales, our RSD analysis will be performed only for scales above a minimum value $s_{\rm min}$, namely $s>s_{\rm min}$ ($s_{\rm min}=32$ \mpc in our baseline settings, see Section \ref{sec:model}). 
Introducing this selection in Equations~\ref{eq:mod2pcf_2} and~\ref{eq:mod2pcf_3},
and noting for clarity $A(\spara') = \bar{n}(\chi_{\rm mod}+\spara'/2)$ and $B = \int_{0}^{\infty} \bar{n}^2(\chi)d\chi$,
we end up with the following modified 2PCF:
\begin{subequations}
\begin{align}
\xi^{\rm mod}_{\rm cut}(\sperp,\spara) = &
\; \xi(\sperp,\spara)\\
& + \int_{{\spara^{\rm min}(\sperp)}<|\spara'|<\spara^{\rm max}}
 (-2 A(\spara') + B) \cdot
 \xi(\sperp,\spara') d\spara',
\end{align}
\label{eq:mod2pcffinal}
\end{subequations}
where $\spara^{\rm min}(\sperp)$ is defined 
as $\spara^{\rm min}(\sperp) =\sqrt{(s_{\rm min}^{\rm cut})^2-\sperp^2}$ with $s_{\rm min}^{\rm cut}$ the minimum value of $s$ 
used in the correction. Except stated otherwise, $s_{\rm min}^{\rm cut}$ is fixed at $s_{\rm min}$, i.e. the minimum scale used in the RSD analysis. The right-column panels of Figure~\ref{fig:2DdataMocks} shows the modified 2PCF defined in Equation~\ref{eq:mod2pcffinal}: though cutting out scales smaller than $s_{\rm min}$ in the integration removes less of the clustering amplitude for $\sperp<s_{\rm min}$ for both data and EZmocks (top and bottom), one can see that the efficiency to reduce angular systematics (the two right panels in last row of Figure~\ref{fig:2DdataMocks}) is of the same order as that of Equation~\ref{eq:mod2pcf}, where no cut is imposed in the integration.

\begin{figure}
\includegraphics[width=0.95\columnwidth]{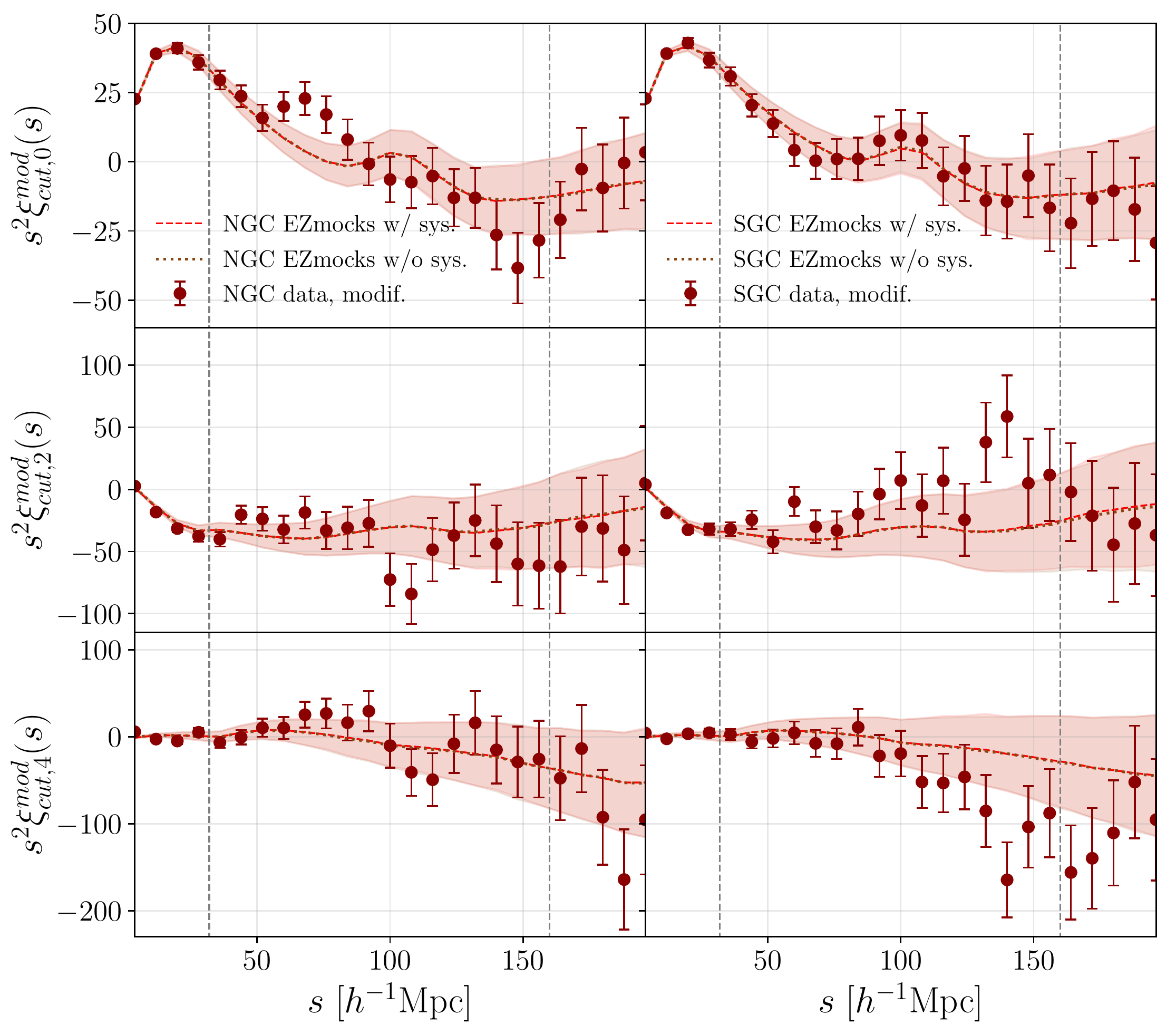}
\caption{\label{fig:datacorrmodif} Multipoles of the modified 2PCF as measured from the eBOSS ELG data sample in each cap and from the mean of the shuffled-z EZmocks with and without systematics. The bands represent the one sigma dispersion around the mean of the mocks. We note that EZmocks with and without systematics mostly overlap, as a result of angular systematics being removed by the modified 2PCF. Errors on data points come from one sigma dispersion of mocks with systematics. Vertical 
dashed lines define 
the baseline fitting 
range.
}
\end{figure}

Equation~\ref{eq:mod2pcffinal} is the modified 2PCF we use in this paper for the RSD analysis for both measurements (on data and mocks) and modelling. We can then define Legendre multipoles $\xi^{\rm mod}_{{\rm cut},\ell}$ using 
Equations~\ref{multipole_def} or 
\ref{eq:effmult}. 
Multipoles of the modified 2PCF with a cut $s_{\rm min}^{\rm cut}=32$\mpc as measured from the eBOSS ELG sample in separate caps and from EZmocks with and without systematics are shown in Figure \ref{fig:datacorrmodif} using $z_{\rm mod}=0.83$ and $s_{\parallel}^{\rm max}=190\Mpch$. 
EZmocks and data are more in agreement than 
in the case of the standard 2PCF multipoles, shown in Figure \ref{fig:datacorr}. It thus suggests that 
removing some of the angular modes allowed us to 
partially remove systematics.

We emphasize that the modified 2PCF introduced in Equation \ref{eq:mod2pcf} does not aim at providing a model for the shuffled 2PCF defined in Equation~\ref{eq:shuffle2PCF}. 
It is a 2PCF estimator that acts similarly to the shuffled 2PCF and removes angular modes significantly.
Our need to discard small scales in the integration over $\spara'$ in Equations~\ref{eq:mod2pcf_2} and \ref{eq:mod2pcf_3}, because of model inaccuracies, led us to adopt Equation \ref{eq:mod2pcffinal} as a final 2PCF estimator, for both measurements and modelling.


\subsection{Reconstruction} \label{sec:reconstruction}

For the isotropic BAO part of the combined RSD+BAO measurements, we use the reconstructed galaxy field to improve our measurements \citep{Eisenstein:2007aa}. Indeed applying reconstruction 
aims at correcting large-scale velocity flow effects, sharpening the BAO peak.

The reconstruction method used in this study follows the works of \cite{Burden:2015aa} and \cite{Bautista:2018aa} which describe a procedure to remove RSD effects. We apply three iterations and assume for the eBOSS ELG sample a linear bias $b=1.4$ and a growth rate $f=0.82$. The smoothing scale is set at $15\Mpch$. \cite{Vargas-Magana:2018aa} showed that the choice of parameter values and cosmology used for reconstruction induces no bias in BAO measurements.

RSD measurements rely on
the pre-reconstruction multipoles and those are then used jointly with the post-reconstruction monopole for the combined RSD+isotropic BAO fit.

\subsection{Covariance matrix}
\label{sec:cov}
We estimate the multipole covariance matrix from the 1000 EZmocks as:
\begin{equation}
\vect{C}_{ij}^{\ell\ell'} = \cfrac{1}{N - 1}\displaystyle \sum_{n = 1}^{N}\left[\xi_{\ell}^n(s_i) - \bar{\xi}_{\ell}(s_i)\right]\left[\xi_{\ell'}^n(s_j) - \bar{\xi}_{\ell'}(s_j)\right]
\end{equation}
where $N$ is the number of EZmocks, $(\ell,\ell')$ are multipole orders, $(i,j)$ run over the separation bins and $\bar{\xi}_{\ell}(s_i)$ is the average value over mocks for multipole $\ell$ in 
bin $s_i$:
\begin{equation}
\bar{\xi}_{\ell}(s_i) = \cfrac{1}{N} \displaystyle \sum_{n = 1}^{N} \xi_{\ell}^n(s_i)
\end{equation}
In the case of RSD fitting, we 
use the first three even Legendre pre-reconstruction multipoles, $\ell=0,2,4$. The procedure is the same whether we use the standard 2PCF or the modified one of Section \ref{sec:modif2PCF}.
In the case of RSD+BAO fitting, we also consider the post-reconstruction monopole, so $\ell=0,2,4,0_{\rm rec}$, where $0_{\rm rec}$ stands for the latter.

We then follow the procedure described in \cite{Hartlab:2007aa} to obtain an unbiased estimator of the inverse covariance matrix, and multiply the inverse covariance matrix from the mocks by a correction factor $(1-(N_d+1)/(N_m-1))$ where $N_m$ is the number of mocks and $N_d$ the number of bins used in the analysis. To account for the uncertainty in the covariance matrix estimate, we rescale the fitted parameter errors 
as proposed in \cite{Percival:2014aa}.

\begin{figure}
\includegraphics[width=0.95\columnwidth]{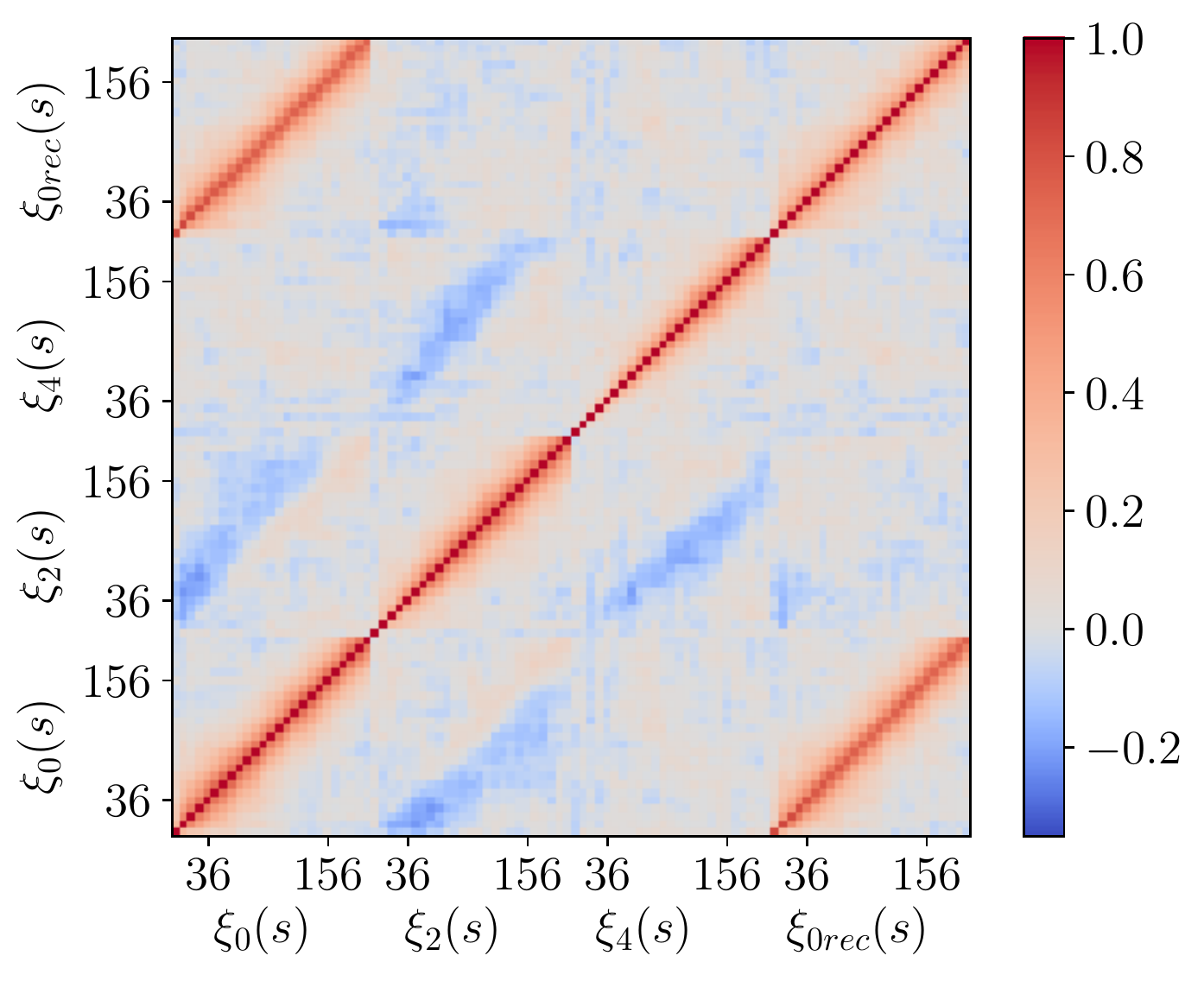}\\
\includegraphics[width=0.95\columnwidth]{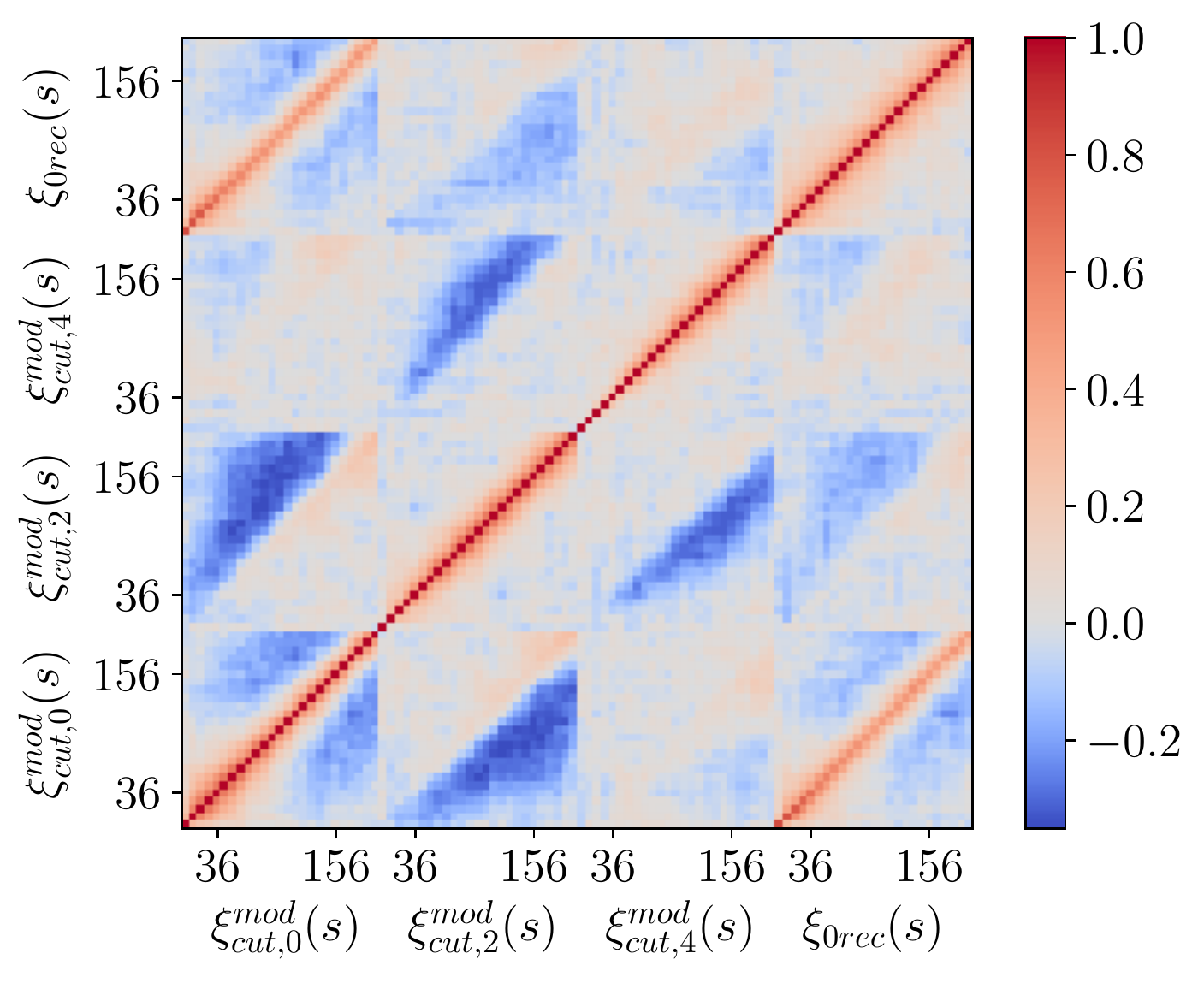}
\caption{\label{fig:corr_matrix} The complete RSD+BAO correlation matrices from 1000 EZmocks computed in 8\mpc bins from 0 to 200$\Mpch$ for  the combined NGC+SGC sample, using the standard (top) and modified (bottom) RSD 2PCF. The latter is computed with $s_{\rm min}^{\rm cut}=32\Mpch$, $z_{\rm mod}=0.83$, $s_{\parallel}^{\rm max}=190$\mpc as in the baseline analysis. The post-reconstruction monopole for BAO is always computed from the standard 2PCF. On both axes we show the fiducial range of the RSD analysis, from 36 to 156\mpc in central bin values.}
\end{figure}

Figure \ref{fig:corr_matrix} shows the correlation matrices computed from the 1000 EZmocks, using the definition $\vect{C}_{ij} / \sqrt{\vect{C}_{ii}\vect{C}_{jj}}$ for the 4 multipoles and their cross-correlations, that are used for the baseline RSD+BAO analysis.
We can notice the differences between the modified and standard 2PCF multipoles, anti-correlations being stronger for the modified 2PCF than for the standard one. On the other hand, the pre- and post-reconstruction monopoles are less strongly correlated when the modified 2PCF is used.

\section{Model}  \label{sec:model}
\subsection{RSD : CLPT-GS model}

Galaxy redshift measurements are a combination of the Hubble rate of expansion and the peculiar velocity of galaxies along the line-of-sight. Therefore what we are effectively measuring is a combination of both the matter density field and the velocity field. The galaxy correlation function is thus affected by multiple sources of non-linearities that are theoretically challenging to model. \cite{Kaiser:1987aa} was the first to derive the linear theory formalism in redshift space, to describe the effect of the peculiar motion of galaxies causing an apparent contraction of the structures along the line-of-sight. \cite{Hamilton:1992aa} then extended the formalism to real space. However the formalism is valid only on scales larger than $\sim 80\Mpch$, where we assume a linear coupling between the matter and velocity fields:
\begin{equation}
\nabla \cdot \vect{v}  = -f\delta_{\rm m},
\end{equation}
where $f$ is the growth rate of structure, $\vect{v}$ the velocity field and $\delta_m$ the underlying matter density field. On smaller scales, the non-linear coupling between the velocity and the matter density fields becomes non-negligible and we need therefore to extend the above formalism beyond linear theory to account for the small-scales non-linearities. \\

In this work, we adopt the same perturbative approach that was previously used in other publications from BOSS \citep{Alam:2015aa, Satpathy:2017aa} and eBOSS \citep{Zarrouk:2018aa,LRG_corr} to model RSD on quasi-linear scales ($\sim 30 - 80$ \mpc), by combining the Lagrangian Perturbation Theory with Gaussian Streaming model.

\subsubsection{CLPT} \label{clpt}
The Convolution Lagrangian Perturbation Theory (CLPT) was introduced by \cite{Carlson:2013aa} to give accurate predictions for correlation functions in real and redshift spaces for biased tracers. In this framework, we perform a perturbative expansion of the displacement field $\vect{\Psi}(\vect{q},t)$.
With this approach, $\vect{\Psi}$ traces the trajectory of a mass element starting from an initial position $\vect{q}$ in Lagrangian coordinates to a final position $\vect{x}$ in Eulerian coordinates through:
\begin{equation}
\vect{x}(\vect{q},t) = \vect{q} + \vect{\Psi}(\vect{q},t),
\end{equation}
where the first order solution of this expansion corresponds to the Zel'dovich approximation \citep{Zeldovich:1970aa, White:2014aa}. Under the assumption that the matter is locally biased, the tracer 
density field, $\delta_X(\vect{x})$, can be written in terms of the Lagrangian bias function $F$ of a linear dark matter field $\delta_{\rm m}(\vect{x})$:
\begin{equation}\label{eq:biasfct}
1+\delta_X(\vect{x}) = F[\delta_{\rm m}(\vect{x})]
\end{equation}
The CLPT model from \citet{Carlson:2013aa} uses contributions up to second order bias, $F_1$ and $F_2$ whose explicit expression can be found in \citet{Matsubara:2008aa}. The first Lagrangian bias $F_1$ is related to Eulerian bias on large scale through $b=1+F_1$.

According to $N$-body simulations \citep{Carlson:2013aa}, the CLPT model performs very well for the real space correlation function down to very small scales (10\mpc). It also shows a good accuracy for the monopole of the correlation function in redshift space down to $\sim 20$ \mpc. However, it suffers from some inaccuracies on quasi-linear scales (30-80\mpc) for the quadrupole in redshift space. To overcome this, \cite{Wang:2014aa} proposed to extend the above formalism by combining it with the Gaussian Streaming Model (GS) proposed by \cite{Reid:2011aa}. The method considers the real space correlation function $\xi(r)$, the pairwise infall velocity $v_{12}(r)$ and the velocity dispersion $\sigma_{12}(r)$ computed from CLPT as inputs to the GS model, as will be described in the next Section. 
The expressions for these functions in the CLPT model are given below (see \citealt{Wang:2014aa} for more details): 
\begin{eqnarray}
1 + \xi(r) &=& \displaystyle \int d^3q M_0(r,q)\\
v_{12}(r) &=& \left[1 + \xi(r)\right]^{-1} \displaystyle \int d^3qM_{\rm 1,n}(r,q)\\
\sigma_{\rm 12,nm}^2(r) &=& \left[1 + \xi(r)\right]^{-1} \displaystyle \int d^3qM_{\rm 2,nm}(r,q)\\
\sigma_\parallel^2(r) &=& \displaystyle \sum_{\rm nm}\sigma_{\rm 12,nm}^2\hat{r}_n\hat{r}_m\\
\sigma_\perp^2(r) &=& \displaystyle \sum_{\rm nm}\left(\sigma_{\rm 12,nm}^2\delta_{\rm nm}^{\rm K} - \sigma_\parallel^2\right)/2
\end{eqnarray}

Here $M_0(r,q)$, $M_{\rm 1,n}(r,q)$ and $M_{\rm 2,nm}(r,q)$ are convolution kernels that depend on a linear matter power spectrum $P_{\rm lin}(k)$ and the first two Lagrangian bias parameters, as the bias expansion is up to second order. The vectors $\hat{r}_n$, $\hat{r}_m$ are unit vectors along the direction of the pair separation, $\sigma_{\rm 12,nm}^2$ is the pairwise velocity dispersion tensor and $\delta_{\rm nm}^{\rm K}$ is the Kronecker delta.
The code\footnote{\url{https://github.com/wll745881210/CLPT_GSRSD}} used in this paper to perform the CLPT calculations was developed by \cite{Wang:2014aa}. 
We use the software \textsc{camb} \citep{Lewis:2000aa} to compute the linear power spectrum $P_{\rm lin}(k)$ for the fiducial cosmology used for the fitting, namely the BOSS cosmology (Equation \ref{eq:cosmoBOSS}), except for the OR mocks.

\subsubsection{The Gaussian Streaming model}
In the GS model, the redshift space correlation function $\xi^s(s_\perp, s_\parallel)$ is modelled as:
\begin{equation} \label{gs_equ}
1 + \xi^s(s_\perp, s_\parallel) = \displaystyle \int dr_\parallel \left[1 + \xi(r)\right] \mathcal{P}(r_\parallel)
\end{equation}
where
\begin{equation}\label{eq:gsclpt}
\mathcal{P}(y) = \cfrac{1}{\sqrt{2\pi}\sigma_{12}(r,\mu)} \exp{\left\lbrace-\cfrac{\left[s_\parallel - r_\parallel - \mu v_{12}(r)\right]^2}{2\sigma_{12}^2(r,\mu)} \right\rbrace}
\end{equation}
and 
\[\sigma_{12}^2(r,\mu) = \mu^2\sigma_\parallel^2(r) + (1-\mu^2)\sigma_\perp^2(r) + \sigma_{\rm FoG}^2\]

$r_\parallel$ corresponds to the line-of-sight separation in real space, while $s_\parallel$ is the line-of-sight separation in redshift space and $s_\perp$ is the transverse separation both in redshift and real spaces. The quantity $r = \sqrt{r_\parallel^2 + s_\perp^2}$ gives the pair separation in real space, and $\mu = r_\parallel/r$ corresponds to the cosine of the angle between the pair separation vector $r$ and the line of sight separation in real space $r_\parallel$. The parameter $\sigma_{\rm FoG}$ accounts for the proper motion of galaxies on small scales \citep{Jackson:1972aa,Reid:2011aa}, causing an elongation of the distribution of galaxies along the line of sight, an effect known as the \textit{Finger of God}. In practice, $\sigma_{\rm FoG}$ is an isotropic velocity dispersion whose role is to account for the scale-dependence of the quadrupole on small scales.

\subsection{Radial Integral Constraint} \label{RICsection}

In this Section, we discuss the impact of the shuffled scheme used for redshift assignment in the random catalogues on the 2PCF measurement and modelling.

The LS estimator from Equation \ref{eq:LS} effectively estimates the observed galaxy correlation function by comparing the observed (weighted) distribution of galaxies to the 3-dimensional survey selection function as sampled by the random catalogue.
In principle, the normalisation of the LS estimator makes it insensitive to the survey selection function, if the random catalogue indeed samples the ensemble average of the galaxy density.
With the shuffled-z scheme, the data radial selection function is directly imprinted on the random catalogue and the density fluctuations are forced to be zero along the line-of-sight: radial modes are suppressed, which effectively modifies clustering measurements on large scales. This so-called radial integral constraint effect is not suppressed by the normalisation of the LS estimator and must be included in the 2PCF modelling. Note that in the case of the eBOSS ELG sample, the impact of the radial selection function is even increased by the division of the survey footprint into smaller chunks accounting for the variations of the radial selection function with imaging depth.

In \cite{de-Mattia:2019aa}, modelling corrections due to the radial integral constraint were derived for the power spectrum analysis. These results are hereafter extended to the correlation function.
The impact of the window function (superscript c) and radial integral constraint (superscript ic) on the correlation function multipoles were modelled in \cite{de-Mattia:2019aa} with the following equation:
\begin{equation}
\xi_{\ell}^{\mathrm{cic}}(s) = \xi_{\ell}^{\mathrm{c}}(s) - IC_{\ell}^{\delta,\mathrm{ic}}(s) - IC_{\ell}^{\mathrm{ic},\delta}(s) + IC_{\ell}^{\mathrm{ic},\mathrm{ic}}(s)
\label{eq:ric}
\end{equation}
where $\xi_{\ell}^{\mathrm{c}}(s)$ are multipoles of the product of the correlation function $\xi$ by the window function (see Equation~2.10 in \cite{de-Mattia:2019aa})
and, for each $(i,j)\in \{(\delta,\mathrm{ic}),(\mathrm{ic},\delta),(\delta,\delta)\}$:
\begin{equation}
IC_{\ell}^{i,j}(s) = \displaystyle \int d\Delta \Delta^{2} \displaystyle\sum_{p}\cfrac{4\pi}{2p+1}\xi_{p}(\Delta)\mathcal{W}_{\ell p}^{i,j}(s,\Delta).
\label{eq:ic}
\end{equation}

$\mathcal{W}_{\ell p}^{i,j}$ are the window function multipoles, as given in equations~2.16 and 2.19 in~\cite{de-Mattia:2019aa}. However, the LS estimator~(Equation \ref{eq:LS}) removes the window function effect with the $RR(s,\mu)$ term in the denominator. Hence, calling $\mathcal{W}_{q}^{\delta,\delta}$ the window function multipoles (e.g. Equation~2.11 in~\cite{de-Mattia:2019aa}), we build the ratios:
\begin{equation}
\mathcal{W}_{\ell p,\mathrm{new}}^{i,j}(s,\Delta) = \frac{2 \ell + 1}{2} \int_{-1}^{1} d\mu \frac{\sum_{q=0}^{q_{\mathrm{max}}} \mathcal{W}_{q p}^{i,j}(s,\Delta) \mathcal{L}_{q}(\mu)}{\sum_{q=0}^{q_{\mathrm{max}}} \mathcal{W}_{q}^{\delta,\delta}(s) \mathcal{L}_{q}(\mu)} \mathcal{L}_{\ell}(\mu)
\end{equation}
to be used instead of the $\mathcal{W}_{\ell p}^{i,j}$ in Equation~\ref{eq:ic}. In practice, we use $q_{\mathrm{max}} = 6$.
In addition, a shot noise contribution to the integral constraint corrections must be accounted for, as given by terms $SN_{\ell}^{ij}(s)$ of Equations~3.6 and 3.7 in \cite{de-Mattia:2019aa}. We proceed similarly to account for the removal of the window function effect in the LS estimator, i.e. instead of the $SN_{\ell}^{ij}(s)$ we use:
\begin{equation}
SN_{\ell,\mathrm{new}}^{i,j}(s) = \frac{2 \ell + 1}{2} \int_{-1}^{1} d\mu \frac{\sum_{q=0}^{q_{\mathrm{max}}} SN_{q}^{i,j}(s) \mathcal{L}_{q}(\mu)}{\sum_{q=0}^{q_{\mathrm{max}}} \mathcal{W}_{q}^{\delta,\delta}(s) \mathcal{L}_{q}(\mu)} \mathcal{L}_{\ell}(\mu).
\end{equation}

In practice, to include the radial integral constraint into our model, we correct the multipoles of the correlation function from the CLPT-GS model, $\xi^{s}$ (as given by Equation \ref{gs_equ}) according to Equation~\ref{eq:ric}.

\subsection{RSD parameter space} \label{param}
We account for the AP effect by introducing two dilation parameters, $\alpha_\perp$ and $\alpha_\parallel$, that rescale the observed separations, $s_\perp, s_\parallel$, into the true ones, $s'_\perp, s'_\parallel$.
Hence, the standard 2PCF model at the true separation is:
\begin{equation}
\xi^s(s'_\perp, s'_\parallel) = \xi^s(\alpha_\perp s_\perp, \alpha_\parallel s_\parallel)
\end{equation}
In our baseline analysis, this $\xi^s(\alpha_\perp s_\perp, \alpha_\parallel s_\parallel)$ is used to compute the radial integral constraint correction (Equation \ref{eq:ric}) and the modified 2PCF (Equation \ref{eq:mod2pcffinal}).

The above dilation parameters relate true values of the Hubble distance $D_{\rm H}(z{\rm eff})$ and 
comoving angular diameter distance $D_{\rm M}(z{\rm eff})$ at the effective redshift to
their fiducial values:
\begin{eqnarray}
\apara &=& \cfrac{D_{\rm H}(z_{\rm eff}) \; r_{\rm drag}^{\rm fid}}{D_{\rm H}^{\rm fid}(z_{\rm eff}) \; r_{\rm drag}}\\
\aperp &=& \cfrac{D_{\rm M}(z_{\rm eff}) \; r_{\rm drag}^{\textnormal{fid}}}{D_{\rm M}^{\textnormal{fid}}(z_{\rm eff}) \; r_{\rm drag}}
\label{eq:alphasAP}
\end{eqnarray}
where the superscript ${\rm fid}$ 
stands for values in the fiducial cosmology and $r_{\rm drag}$ is the comoving sound horizon at the redshift at which the baryon-drag optical depth equals unity \citep{Hu:1996aa}.

The growth rate of structure $f(z)$ defined in Equation \ref{eq:fz} is taken into account in the 
correlation function model via $v_{12}(r)$ and $\sigma_{12}(r)$, as those are proportional to $f(z)$ and $f^2(z)$, respectively. The two Lagrangian biases $F_1$ and $F_2$ as described by Equation \ref{eq:biasfct} are free parameters of the model.
The second Lagrangian bias $F_2$ impacts mainly the small scales \citep{Wang:2014aa} and thus is mostly degenerate with $\sigma_{\rm FoG}$ and not well constrained by the data. Due to its small impact on the scales of interest, we chose to fix $F_2=F_2(F_1)$ using the peak background splitting assumption \citep{Cole:1989aa} with a Sheth-Tormen mass function \citep{Sheth:1999aa}.

Altogether, we thus explore a five dimensional parameter space $\vect{p} = \left\{\alpha_\parallel, \alpha_\perp, f(z), F_1, \sigma_{FoG}\right\}$ in our RSD analysis.
The growth rate and biases being degenerate with $\sigma_8$, we hereafter report values of $f\sigma_8$ and $b_1\sigma_8$, where $b_1=1+F_1$. 
As explained in \citet{gil-marin20a}, to remove the $h$ dependency of $\sigma_8$, we rescale $\sigma_8$ by taking the amplitude of the power spectrum at $8\times\alpha_{\rm iso}$\mpc where $\alpha_{\rm iso}$ is defined hereafter.

\subsection{Isotropic BAO}
An alternative way to parametrize the AP effect is to decompose the distortion into an isotropic and anisotropic shifts.
The isotropic component $\alpha_{\rm iso}$ is related to parallel and transverse shifts, $\alpha_\parallel$ and $\alpha_\perp$, via:
\begin{equation}\label{eq:alphaiso}
\alpha_\mathrm{iso} = \alpha_\parallel^{1/3}\alpha_\perp^{2/3}
\end{equation}
It corresponds to the isotropic shift of the BAO peak position in the monopole of the correlation function; the anisotropic shift $\epsilon$ is defined as $1+\epsilon = \alpha_\parallel\alpha_\perp^{-1/3}$.

BAO measurements from the eBOSS ELG sample in configuration space are presented in \cite{raichoor20a}. 
We hereafter fit the post-reconstruction BAO using the same BAO model as in \cite{raichoor20a}:
\begin{equation}
\xi_\mathrm{BAO}(s, \alpha_\mathrm{iso}) =
B\xi_\mathrm{temp}(\alpha_{\rm iso} \cdot s) + A_0 + A_1/s + A_2/s^2
\end{equation}
where $B$ is the post-reconstruction bias, the $A_i$'s are broadband parameters with $i=0,1,2$. The template $\xi_{\rm temp}$ is the Fourier transform of the following power spectrum:
\begin{equation}
P(k,\mu) = \frac{1+\mu^2\beta e^{-k^2\Sigma_r^2/2}}{1+k^2\mu^2\Sigma_s^2/2}\left(\frac{P_{\rm lin}-P_{\rm nw}}{e^{k^2((1-\mu^2)\Sigma^2_\perp+\mu^2\Sigma_\parallel^2)/2}}+P_{\rm nw} \right)
\end{equation}
where $P_{\rm lin}$ is a linear power spectrum taken from \textsc{camb} and $P_{\rm nw}$ is a 'no-wiggle' power spectrum computed with the formula from \cite{Eisenstein:1998aa}.
We use the same smoothing scales as in \citet{raichoor20a}, i.e. $\Sigma_r=15\Mpch$, $\Sigma_s=3\Mpch$, $\Sigma_\perp=3\Mpch$, $\Sigma_\parallel$=5\mpc and we set $\beta$=0.593 \citep[see also][]{Ross:2016aa, Seo:2016aa}.

For the modelling of the post-reconstruction BAO signal, we do not 
include a radial integral constraint correction as the effect on the post-reconstruction monopole is absorbed by the broadband parameters.

\subsection{Parameter estimation}\label{sec:likelihood}
In this paper, we perform RSD measurements and a joint fit of RSD and isotropic BAO. For both RSD and combined RSD+BAO fits, we use a nested sampling algorithm called \textsc{MultiNest} \citep{Feroz:2009aa} to infer the posterior distributions of the set of cosmological parameters $\vect{p}$. \textsc{MultiNest} is a Monte Carlo method that efficiently computes the Bayesian evidence, but also accurately produces posterior inferences as a by-product. Our analysis makes use of the publicly available python version\footnote{\url{https://johannesbuchner.github.io/PyMultiNest/}} of \textsc{MultiNest}. For the frequentist fits of our analysis, we use the \textsc{minuit} algorithm\footnote{\url{https://github.com/scikit-hep/iminuit}} \citep{James:1975aa} which is specifically used to get the best fits of data and single mocks. The likelihood $\mathcal{L}$ is computed from the $\chi^2$ assuming a Gaussian distribution:
\begin{equation}
\mathcal{L} \propto \exp \left(-\cfrac{\chi^2(\vect{p})}{2}\right)
\end{equation}
where the $\chi^2$ is constructed from the correlation function multipoles measured from data catalogs, $\xi^d_{\ell}$, and predicted by the model, $\xi^{m}_{\ell}$, as follows: 
\begin{equation}
\chi^2(\vect{p}) = \displaystyle \sum_{i,j}^{\ell,\ell'} \left[\xi^d_{\ell}(s_i) - \xi^{m}_{\ell}(s_i,\vect{p})\right]
\vect{C}_{ij}^{\ell\ell'} \left[\xi^d_{\ell'}(s_j) - \xi^{m}_{\ell'}(s_j,\vect{p})\right]
\end{equation}
Here indexes $(i$,$j)$ run over the separation bins and $\vect{C}_{ij}^{\ell\ell'}$ is the inverse covariance matrix computed from the 1000 EZmocks (see Section~\ref{sec:cov}). Indexes $(\ell$,$\ell')$ run over the multipoles of the correlation function, where $\ell=0,2,4$ if RSD only and $\ell=0,2,4,0_{\rm rec}$ if a combined RSD+BAO fit is performed. We recall that $\xi^d_{\ell}$ and $\xi^{m}_{\ell}$ can be computed from a standard 2PCF or a modified one. 
The priors on the parameters of the RSD model are flat priors given in Table \ref{prior}. Performing the joined fit RSD+BAO by combining the likelihoods allows the Gaussian assumption required to combine RSD and BAO posteriors as in \citet{LRG_corr} to be relaxed.

For the RSD+BAO fit, the BAO isotropic shift, $\alpha_{\rm iso}$ is related to the two anisotropic AP parameters through Equation \ref{eq:alphaiso}. However we add an additional prior constraint by adding a flat prior on $\alpha_{\rm iso}$ from 0.8 to 1.2.  Due to reconstruction, the $B$ bias can be different than $b_1$, therefore $B$ is not fixed at $1+F_1$ but is kept as a free parameter. As in \cite{raichoor20a}, we use a Gaussian prior on $B$ of $0.4$ width, centred around the $B$ value obtained from the first bin of the fitting range when setting $A_i$ to zero.

When fitting onto the combined data sample, we chose to have only one set of biases for the whole sample, neglecting the difference between caps.

Unless otherwise specified, we fit the RSD multipoles over a range in separation from 32 to 160 \mpc and from 50 to 150 \mpc for the post-reconstruction monopole, using in both cases 8\mpc bins and the BOSS cosmology (Equation \ref{eq:cosmoBOSS}) at the effective redshifts quoted in Table \ref{ELG_stats}.

\begin{table}
	\centering
	\begin{tabular}{|c|c|c}
		\hline
		Parameter & Min value & Max value \\
		\hline 
		$\alpha_\parallel$, $\alpha_\perp$ & 0.6 & 1.4  \\ 
		$f(z)$ & 0 & 1.5  \\ 
		$F_1$ & -0.2 & 2 \\ 
		$\sigma_{\rm FoG}$ & 0 & 10  \\ 
		\hline 
		$\alpha_{\rm iso}$ & 0.8 & 1.2  \\ 
		$B$ & \multicolumn{2}{c}{Gaussian}\\
		\hline
	\end{tabular}
	\caption{\label{prior} Flat priors on the RSD model parameters used in the cosmological analysis of this paper. }
\end{table}

\section{Tests on Mocks}  \label{sec:analysis}
In this Section, we present tests on mocks in order to validate our analysis. We first demonstrate the robustness of our CLPT-GS model with accurate N-body mocks; we then validate our analysis choices with the approximate EZmocks for both RSD and RSD+BAO fits. Results from the latter tests are presented in Table \ref{singleEZmockfit}.

\subsection{CLPT-GS model validation}
We quantify here the ability of the CLPT-GS model to recover the cosmological parameters from accurate mocks made from the {\sc Outer Rim} N-body simulation. We present a summary of the results, and refer the reader to \citet{Alam2020}, where those are presented in details.

First, the non-blind mocks described in Section \ref{nbody} were analysed. The statistical uncertainty on the recovered parameter values in these accurate mocks are~0.5-0.6\%, 0.3-0.5\% and 1-2\%  in  $\alpha_{\parallel}$, $\alpha_{\bot}$ and $f\sigma_8$, respectively. No statistically significant bias in the parameter values was observed, despite the wide range of ELG HOD models used.

A set of blind mocks was then analysed, to test for possible biases, primarily in $f\sigma_8$, that could arise due to theoretical approximations in the model. To create these mocks, the peculiar velocities of the galaxies were scaled by an undisclosed factor leading to a change in the expected value of $f$ and thus $f\sigma_8$. The other cosmological parameters were unaffected.
The mean deviations of the fitted cosmological parameters with respect to expectations
are 0.9\%, 0.7\% and 1.6\% in $\alpha_{\parallel}$, $\alpha_{\bot}$ and $f\sigma_8$,
showing that the CLPT-GS model 
describes the blind mock catalogues remarkably well. 

These tests on N-body mocks demonstrate that the CLPT-GS model provides unbiased RSD measurements, within the statistical error of the mocks. Following \citet{Alam2020}, we adopt as our modelling systematic errors:  1.8\%, 1.4\% and 3.2\% for  $\alpha_{\parallel}$, $\alpha_{\bot}$ and $f\sigma_8$ respectively. We note that these errors are an order of magnitude smaller than the statistical error of the eBOSS ELG sample (see Section~\ref{sec:results}), and will marginally affect the precision of our measurements.

\subsection{Radial integral constraint modelling}

\begin{figure}
\centering
\includegraphics[width=0.95\columnwidth]{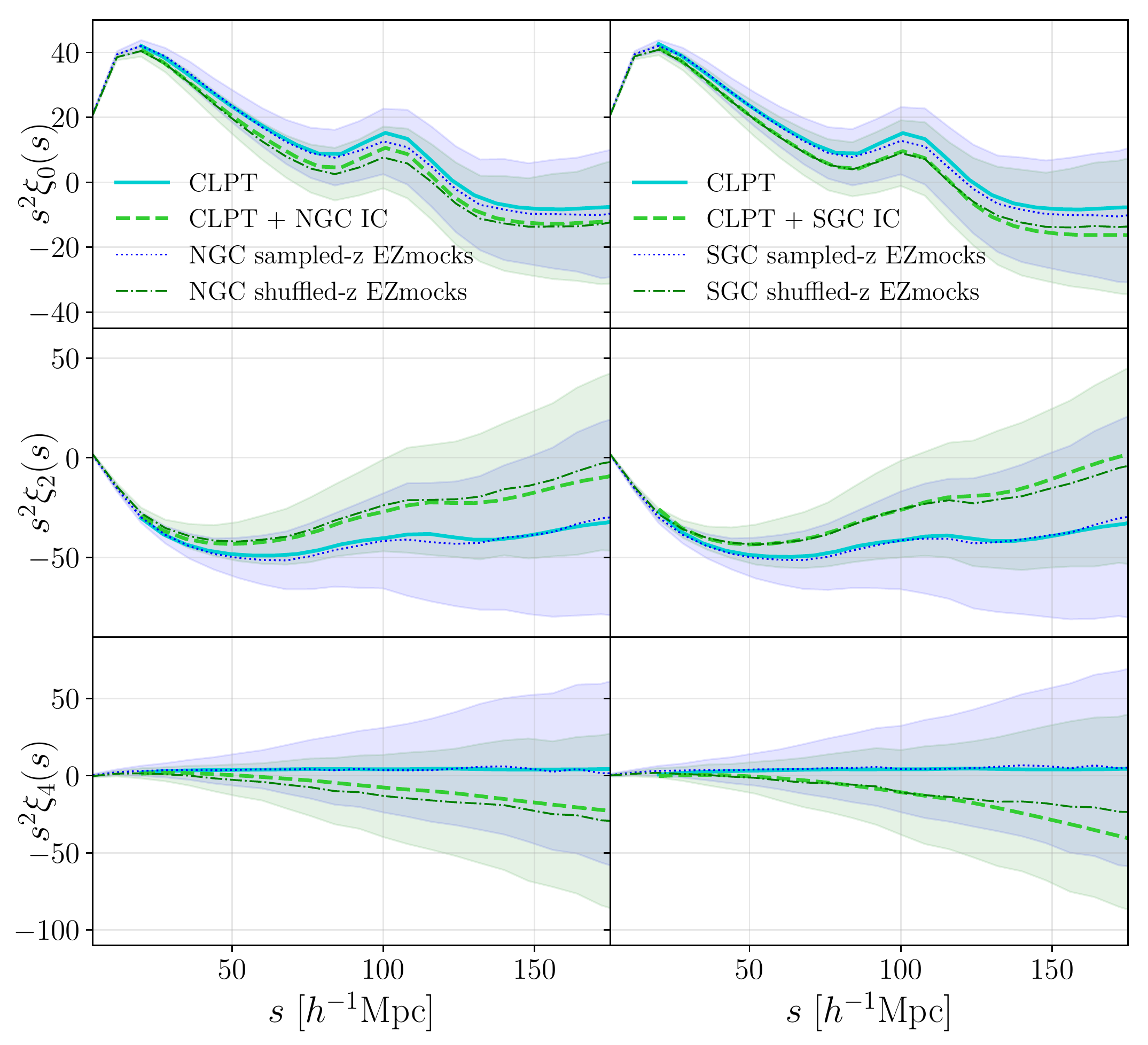}
\caption{Multipoles of the standard 2PCF as measured in EZmocks without systematics, using the 'sampled-z' (blue) or the 'shuffled-z' (green) scheme. Left panels are for the NGC, right panels for the SGC. The dotted lines and the shaded area represent the mean of the mocks and the dispersion at 1$\sigma$ around the mean, respectively. Blue solid and dashed green lines are the CLPT-GS model prediction without and with the radial integral constraint correction, respectively.
Cosmological parameters used for the model are: ($\alpha_{\parallel}$, $\alpha_{\perp}$, $f$, $F_1$, $\sigma_{\rm FoG}$)\; = \; (1.0,1.0,0.84,0.4,2.0).
}\label{fig:RICEZ}
\end{figure}

In Section~\ref{RICsection} we justified the use of the 'shuffled-z' scheme to assign redshifts to random objects, for both data and mocks, in order to reproduce the radial selection function of the survey.
This scheme  has a significant impact on the multipole measurements in the eBOSS ELG sample, as illustrated in Figure~\ref{fig:RICEZ} with the EZmocks (dashed lines with shaded regions), using either the 'sampled-z' scheme (blue) or the 'shuffled-z' one (green). At large scales, the increase in the quadrupole and decrease in the hexadecapole are noticeable. We account for this radial integral constraint effect in our modelling with the formalism presented in Section~\ref{RICsection}, and we test hereafter the impact of that correction on the estimated cosmological parameters. Results are presented in the upper part of Table~\ref{singleEZmockfit}.

The baseline for this test is provided by fits on the 'sampled-z' EZmocks, using a standard 2PCF model based on CLPT-GS at the data effective redshift. When compared to the values expected for our fiducial cosmology, the results show deviations of 2.4\%, 0.2\%, and 0.9\% for $\apara$, $\aperp$, and $\fsig$, respectively. Those small deviations may come from the fact that EZmocks are approximate mocks meant to determine the covariance matrix to be used in the measurements. The linear scales around the BAO are well reproduced but the small scales and hence the full shape fits are not accurate enough for model validation. In this sense, we note that the corresponding value of the isotropic BAO scale $\alpha_{\rm iso}=1.007$ is consistent with the value measured in \citet{raichoor20a} from the post-reconstruction monopole.

Performing a similar fit, i.e. without RIC correction, using the 'shuffled-z' scheme instead of the 'sampled-z' one, the previous cosmological parameter estimations are shifted by 4.0\%, 5.3\% and 4.2\% for $\apara$, $\aperp$ and $\fsig$. Those shifts are large, and explained by the significant differences in the multipoles between the 'sampled-z' and the 'shuffled-z' schemes due to the radial integral constraint effect (Figure~\ref{fig:RICEZ}). It justifies that we correct our modelling for this effect. Including the correction as described in Section~\ref{RICsection}, the deviations are significantly reduced to 2.1\%, 1.2\%, and 0.2\% for $\apara$, $\aperp$, and $\fsig$, respectively.
The growth rate is almost perfectly recovered and the remaining biases in $\alpha_{\parallel}$ and $\alpha_{\perp}$ are reasonable. The observed shifts are taken as systematic errors due to the radial integral constraint (RIC) modelling in our final error budget (see Table~\ref{tab:budget}).

\subsection{Mitigating unknown angular systematics in RSD fits}\label{sec:mitigatesys}

As already mentioned, the eBOSS ELG sample suffers from unknown angular systematics that are not corrected by the photometric weights. These systematic effects bias our cosmological results (see below). In this Section, we show that the modified 2PCF (Section~\ref{sec:modif2PCF}) is efficient at reducing those biases.

Here, our reference consists in fitting a RIC-corrected model onto 'shuffled-z' EZmocks without systematics using the standard 2PCF (see Standard 2PCF, 'no systematics' row in Table~\ref{singleEZmockfit}). Performing a similar fit on the 'shuffled-z' EZmocks \textit{with} systematics, shifts those reference values by 0.3\%, 2.2\% and 9.6\% for $\apara$, $\aperp$, and $\fsig$, respectively. The shift in $\fsig$ is significant and justifies our use of the modified 2PCF defined by Equation~\ref{eq:mod2pcffinal} in Section~\ref{sec:modif2PCF} to cancel the angular modes. 

\begin{figure}
\centering
\begin{tabular}{cc}
\includegraphics[width=0.95\columnwidth]{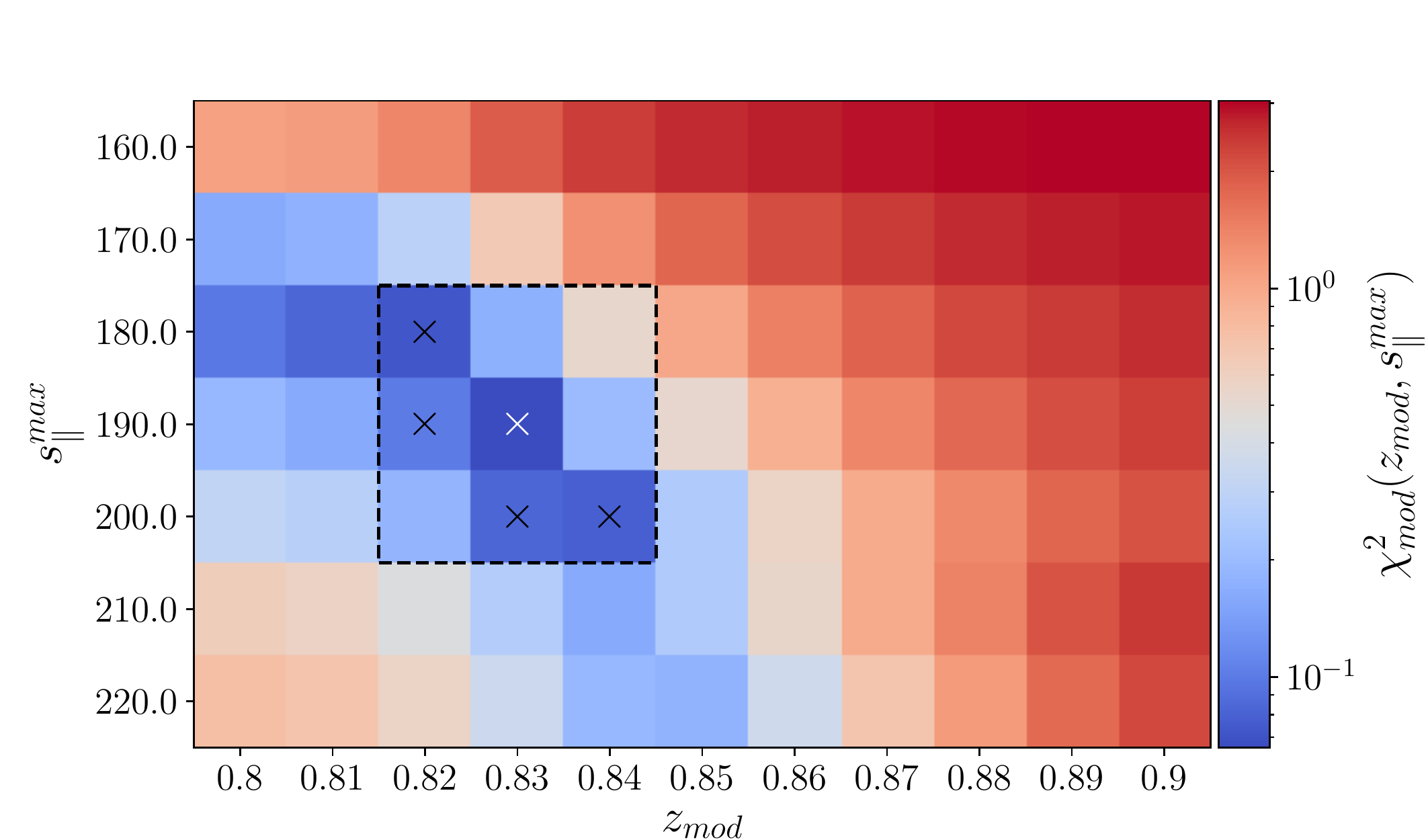} &
\end{tabular}
\caption{\label{fig:chizRPfig} $\chi^2_{\rm mod}$ measuring the difference between multipoles of the modified 2PCF obtained on 'shuffled-z' EZmocks with and without systematics (Equation \ref{eq:chi2zRP}), as a function of the modified 2PCF free parameters, $z_{\rm mod}$ and $\spara^{\rm max}$. The dashed line encompasses the 8 neighbouring pixels around the minimum, marked with a white cross. Black crosses indicate pixels with $\chi^2_{\rm mod}<0.1$. We note that the standard 2PCF would provide $\chi^2 \sim 3$. 
}
\end{figure}

The free parameters $z_{\rm mod}$ and $s_\parallel^{\rm max}$ of the modified 2PCF are chosen by minimising the following quantity:
\begin{equation}\label{eq:chi2zRP}
\chi^2_{\rm mod}(z_{\rm mod},s_\parallel^{\rm max}) = \Delta \; (C^{\rm mod}_{\rm cut})^{-1} \;  \Delta^{\rm T},
\end{equation}
where $\Delta=\xi^{\rm mod}_{\rm cut,syst}(z_{\rm mod},s_\parallel^{\rm max}) - \xi^{\rm mod}_{\rm cut}(z_{\rm mod},s_\parallel^{\rm max})$ is the vector of differences between the multipoles of the modified 2PCF ($\ell=0,2,4$) measured from the mean of the EZmocks with and without systematics and restricted to our fiducial fitting range in $s$, and $C^{\rm mod}_{\rm cut}$ is the covariance matrix built from the 1000 EZmocks without systematics, using the modified 2PCF. The minimisation yields $z_{\rm mod}=0.83$ and $s_\parallel^{\rm max}=190\Mpch$. In the following we will choose those two parameters as our baseline choice.
The 2D variations of $\chi^2_{\rm mod}$ with respect to both parameters are represented in Figure~\ref{fig:chizRPfig}, which shows a valley around our minimum (represented by the darker blue pixels). 
The minimum is well defined at the center of this valley.
Moreover, the minimum $\chi^2_{\rm mod}$ reaches a value below 0.1 that indicates that the modified 2PCF successfully mitigates the systematic effects introduced in the mocks. Using the covariance matrix with systematics or using the modified 2PCF with no cut in $s$ (see Equation~\ref{eq:mod2pcf}) result in the same minima.

To quantify the systematic error related to the modified 2PCF, we compare in Table~\ref{singleEZmockfit} fit results to the modified and standard 2PCF multipoles from the mean of the mocks for the 'no systematics' case (we recall that we use 'shuffled-z' EZmocks and a RIC-corrected model as baseline). We find deviations of 0.3$\%$, 0.04$\%$, 1.4$\%$ in $\alpha_{\parallel}$, $\alpha_{\perp}$ and $f\sigma_8$, respectively. Then, we vary $z_{\rm mod}$ and $ s_\parallel^{\rm max}$ around their nominal values and take as a systematic error the largest of the observed deviations for each parameter. For $z_{\rm mod}$, we obtain 0.3$\%$, 0.2$\%$ and 1.8$\%$ for $\alpha_{\parallel}$, $\alpha_{\perp}$ and $f\sigma_8$, respectively. For $s_\parallel^{\rm max}$, the equivalent numbers are 0.2$\%$, 0.3$\%$ and 1.5$\%$. The error we assign to using the modified 2PCF in the absence of systematics is taken conservatively as the sum in quadrature of the three effects previously described, which amounts to 0.5$\%$, 0.4$\%$, 2.7$\%$ for $\alpha_{\parallel}$, $\alpha_{\perp}$ and $f\sigma_8$, respectively. We also show that taking more extreme values for the parameters ($s_\parallel^{\rm max}=200$\mpc, $z_{\rm mod}=0.87$ and $s_\parallel^{\rm max}=100$\mpc, $z_{\rm mod}=0.84$) implies deviations that are at same level. This shows the robustness of the modified 2PCF to recover the correct values of the cosmological parameters in the absence of systematic effects in the mocks.
We note that larger biases are observed when the parameter $s_\parallel^{\rm min}$ of the modified 2PCF is set to 0 in Equation \ref{eq:mod2pcffinal} (see 'no cut' label in Table~\ref{singleEZmockfit}). This especially the case for $f\sigma_8$, which is expected since using the model for very small scales, where it is invalid, distributes model inaccuracies over all scales.

We now study the response of the modified 2PCF in the case of shuffled-z EZmocks with 'all systematics' (and a RIC-corrected model). Deviations with respect to results from the modified 2PCF and mocks without systematics are 0.3$\%$, 0.9$\%$ and 1.7$\%$ for $\alpha_{\parallel}$, $\alpha_{\perp}$ and $f\sigma_8$ respectively, showing a significant reduction with respect to the corresponding results from the standard 2PCF reported at the beginning of the Section. This demonstrates that the modified 2PCF is key for this analysis as it reduces the bias on $f\sigma_8$ by a factor of nearly 6. When compared to results from the standard 2PCF and mocks without systematics, the deviations in cosmological parameters are small (0.7$\%$, 0.9$\%$ and 0.4$\%$ for $\alpha_{\parallel}$, $\alpha_{\perp}$ and $f\sigma_8$), nevertheless, there is a mild increase of the dispersion of about 10$\%$ for $\alpha_{\parallel}$, 15$\%$ for $\alpha_{\perp}$, and 15$\%$ for $f\sigma_8$. 

We also evaluate the impact of changes of $z_{\rm mod}$ and $s_\parallel^{\rm max}$ around their nominal values by considering the 8 neighbouring pixels around the minimum defined in Figure~\ref{fig:chizRPfig}, 
which correspond to changes 
of $\Delta z_{\rm mod}\in\{0,\pm0.01\}$ and $\Delta \spara^{\rm max}\in\{0,\pm10\}\Mpch$.
First, we consider the shifts induced by a small increase in $\chi_{\rm mod}^{2}$, i.e. pixels marked by black crosses in Figure~\ref{fig:chizRPfig} which have $\chi_{\rm mod}^{2}<0.1$. The largest deviations with respect to mocks without systematics and 
the modified 2PCF with baseline parameters are obtained for $s_\parallel^{\rm max}=200\Mpch$ and $z_{\rm mod}=0.84$: 0.3$\%$, 0.9$\%$ and 2.1$\%$ for $\alpha_{\parallel}$, $\alpha_{\perp}$ and $f\sigma_8$ respectively. These numbers become 0.6$\%$, 0.8$\%$ and 0.7$\%$ when the comparison is made w.r.t. the standard 2PCF.  
The deviations are only marginally larger than those previously quoted, as expected since we are close to the minimum.
Considering all neighbouring pixels, the largest biases are obtained for $s_\parallel^{\rm max}=180\Mpch$ and $z_{\rm mod}=0.84$, which  is the neighbouring pixel with the largest $\chi_{\rm mod}^2$ value.
With respect to mocks without systematics and the modified 2PCF with baseline parameters, we observe deviations of 0.4$\%$, 1.3$\%$ and 3.6$\%$ for $\alpha_{\parallel}$, $\alpha_{\perp}$ and $f\sigma_8$.
These numbers become 0.7$\%$, 1.2$\%$ and 2.3$\%$ when comparing to the standard 2PCF case.
In the case of $f\sigma_8$, this is about twice the deviation observed for our baseline parameters when using the modified 2PCF and six times with the standard 2PCF.
While the cosmological parameters are still better recovered with the modified 2PCF than with the standard one, the above results underline that the mitigation efficiency of the modified 2PCF strongly depends on the values of its two free paremeters.
For completeness, we observe that  settings $s_\parallel^{\rm max}$ or $z_{\rm mod}$ to more extreme values (100$\Mpch$ and 0.87, respectively) degrades significantly the efficiency of mitigation: this is understood since the systematics are no longer corrected as efficiently as with the baseline parameters.

\begin{figure}
\centering
\includegraphics[width=0.95\columnwidth]{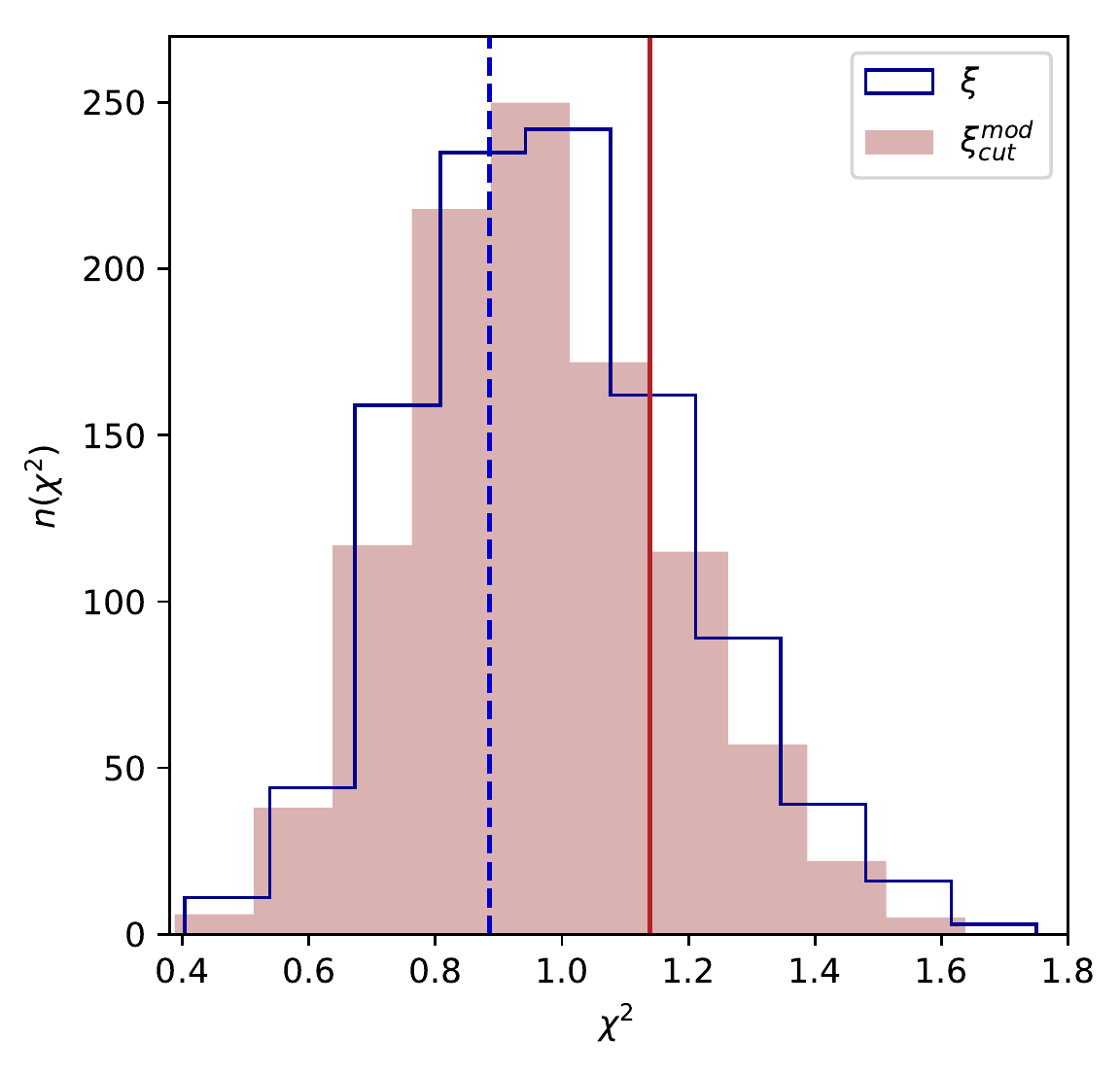}
\caption{\label{fig:EZmockschi2} $\chi^2$ distributions for RSD fits to multipoles of the standard (step line in blue) and modified 2PCF (filled in red) (using a cut in $s$ at 32 \mpc for the latter) in 1000 EZmocks with systematics. Dashed blue and solid red vertical lines are the $\chi^2$ for the eBOSS ELG data sample for the standard and modified 2PCF, respectively.
}
\end{figure}

\begin{figure*}
\centering
\includegraphics[width=1.0\textwidth]{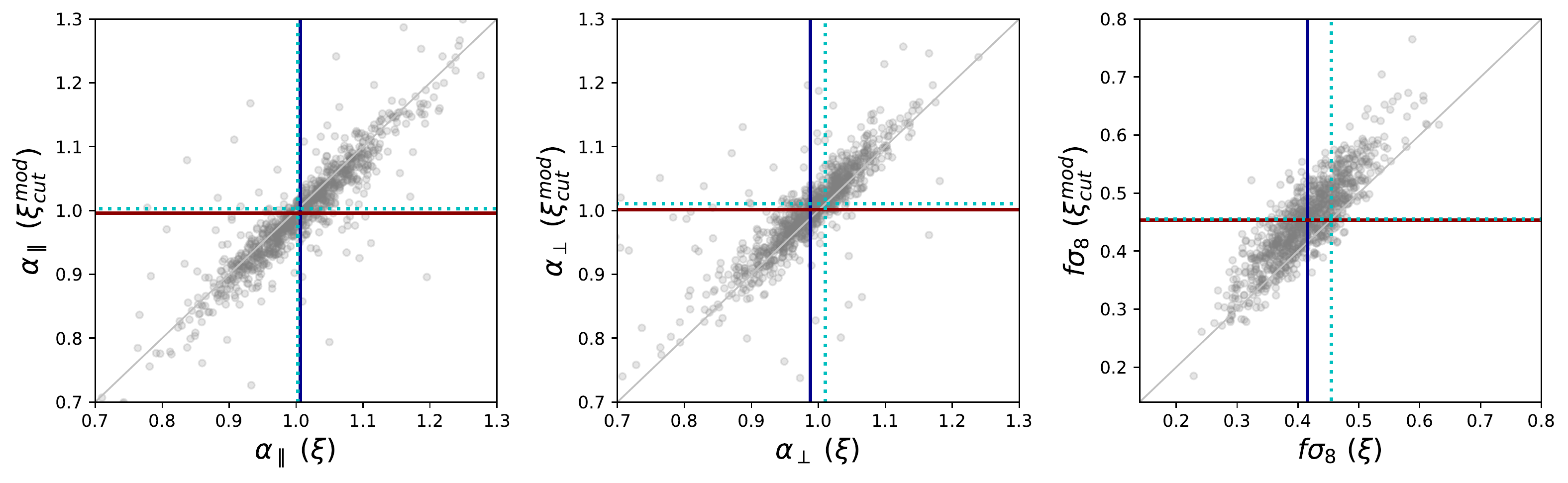}
\caption{\label{fig:EZmocksComp2PCF} Comparison of parameter best-fit values for RSD fits based on the standard 2PCF (x-axis) and the modified one (y-axis) obtained for the 1000 EZmocks with systematics. Cyan dotted lines correspond to the mean of the fits on EZmocks without systematics using the standard 2PCF, which is the reference case here. Solid lines indicate the mean values of the parameters for the modified (red horizontal) and standard (blue vertical) 2PCF. The solid gray line is the identity. 
}
\end{figure*}

Results of fits to EZmocks with systematics using the modified 2PCF (with ($s_\parallel^{\rm max}$, $z_{\rm mod}$) at their baseline values) and the standard one are compared in Figures~\ref{fig:EZmockschi2} and~\ref{fig:EZmocksComp2PCF}. Both the standard and modified 2PCFs provide similar $\chi^2$ distributions, but due to systematics, the standard 2PCF fits are driven by extra-correlations in the quadrupole at intermediate scales (see middle panels of Figure~\ref{fig:datacorr}) which results in clearly biased values for  $\alpha_{\perp}$ and $f\sigma_8$.
Figure~\ref{fig:EZmocksComp2PCF} shows that on average, the modified 2PCF brings a significant improvement for these two parameters.

\subsection{Joined RSD+BAO fit}
As in \cite{demattia20a}, we perform a joined fit of RSD and isotropic BAO. We take into account the cross-correlation between the pre-reconstruction multipoles and the post-reconstruction monopole, and combine their likelihoods, as explained in Section \ref{sec:likelihood}.

When fitting the 'shuffled-z' EZmocks without systematics using the standard 2PCF, combining with isotropic BAO has a small effect on the median best-fit parameter values of individual mocks. We indeed observe shifts of 0.2$\%$, 0.2$\%$ and 0.9$\%$ for $\alpha_{\parallel}$, $\alpha_{\perp}$ and $f\sigma_8$, respectively (see second part of Table \ref{singleEZmockfit}). The same is observed when using the modified 2PCF with the baseline parameters in the RSD part of the fit: shifts are of 0.3$\%$, 0.2$\%$ and 0.2$\%$ for $\alpha_{\parallel}$, $\alpha_{\perp}$ and $f\sigma_8$ compared to pure RSD fits with the modified 2PCF.

As already observed for pure RSD fits, adding systematics biases a lot the results compared to fits on EZmocks without systematics; for RSD+BAO fits with the standard 2PCF, all parameters are biased low, by 2.6$\%$ for $\alpha_{\parallel}$,  4.2$\%$ for $\alpha_{\perp}$ and 8.8$\%$ for $f\sigma_8$. For the AP parameters, these deviations are larger than in the RSD fits. 

It again motivates the use of the modified 2PCF to mitigate the systematics. As compared to RSD+BAO fits on mocks without systematics  using the modified (standard) 2PCF, RSD+BAO fits with the modified 2PCF on mocks with systematics deviate by only 0.1$\%$ (0.9$\%$) for $\alpha_{\parallel}$, 0.6$\%$ (0.8$\%$) for $\alpha_{\perp}$ and 1.9$\%$ (1.1$\%$) for $f\sigma_8$, which are comparable to those in the pure RSD case. This suggests that with the standard 2PCF, RSD+BAO fits are driven by systematics in pre- and post-reconstruction multipoles that can be correlated and highlights again the need for the modified 2PCF.

\renewcommand{\arraystretch}{1.1}
\begin{table*}
\centering
\begin{tabular}{|l@{\hspace{2cm}}|c@{\qquad}|c@{\qquad}|c@{\qquad}|}
\hline 
 & $\alpha_{\parallel}$ & $\alpha_{\perp}$ & $\quad f\sigma_8$ \\ 
\hline 
\hline
RSD & \\
\hline
& Standard 2PCF \\
\hline
no systematics \& sampled-z, no IC corrections  & $1.024^{+0.069}_{-0.071}$ & $0.998^{+0.053}_{-0.054}$ & $0.454^{+0.057}_{-0.058}$ \\
no systematics, no IC corrections & $0.983^{+0.065}_{-0.068}$ & $1.051^{+0.057}_{-0.059}$ & $0.435^{+0.055}_{-0.055}$ \\ 
{\bf no systematics} & \boldmath$1.003^{+0.067}_{-0.067}$ & \boldmath$1.010^{+0.053}_{-0.055}$ & \boldmath$0.455^{+0.057}_{-0.055}$ \\ 
all systematics & $1.006^{+0.075}_{-0.073}$ & $0.988^{+0.058}_{-0.056}$ & $0.415^{+0.056}_{-0.057}$\\
\hline
& Modified 2PCF \\
\hline
{\bf no systematics, baseline (\boldmath$\spara^{\rm max}=190$\mpc, \boldmath$z_{\rm mod}=0.83$}) & \boldmath$0.999^{+0.067}_{-0.070}$ & \boldmath$1.011^{+0.053}_{-0.056}$ & \boldmath$0.462^{+0.066}_{-0.067}$ \\
no systematics, $\spara^{\rm max}=100$\mpc, $z_{\rm mod}=0.84$ & $0.996^{+0.070}_{-0.064}$ & $1.011^{+0.054}_{-0.048}$ & $0.462^{+0.060}_{-0.056}$ \\
no systematics, $\spara^{\rm max}=180$\mpc, $z_{\rm mod}=0.83$ & $1.004^{+0.066}_{-0.070}$ & $1.015^{+0.054}_{-0.058}$ & $0.461^{+0.062}_{-0.055}$ \\
no systematics, $\spara^{\rm max}=190$\mpc, $z_{\rm mod}=0.82$ & $0.999^{+0.067}_{-0.070}$ & $1.012^{+0.054}_{-0.055}$  & $0.464^{+0.071}_{-0.073}$ \\
no systematics, $\spara^{\rm max}=190$\mpc, $z_{\rm mod}=0.84$ & $1.001^{+0.066}_{-0.067}$ & $1.011^{+0.054}_{-0.054}$ & $0.460^{+0.065}_{-0.061}$ \\
no systematics, $\spara^{\rm max}=200$\mpc, $z_{\rm mod}=0.83$ & $1.001^{+0.063}_{-0.069}$ & $1.013^{+0.056}_{-0.055}$ & $0.462^{+0.068}_{-0.069}$ \\
no systematics, $\spara^{\rm max}=200$\mpc, $z_{\rm mod}=0.87$ & $1.002^{+0.066}_{-0.071}$ & $1.012^{+0.053}_{-0.055}$ & $0.459^{+0.057}_{-0.057}$ \\
no systematics, no cut & $1.020^{+0.088}_{-0.082}$ & $1.012^{+0.055}_{-0.063}$ &  $0.435^{+0.112}_{-0.110}$ \\
\hline
{\bf all systematics, baseline (\boldmath$\spara^{\rm max}=190$\mpc, \boldmath$z_{\rm mod}=0.83$}) & \boldmath$0.996^{+0.075}_{-0.075}$ & \boldmath$1.001^{+0.063}_{-0.062}$ & \boldmath$0.454^{+0.066}_{-0.065}$ \\
all systematics, $\spara^{\rm max}=100$\mpc, $z_{\rm mod}=0.84$ & $0.995^{+0.073}_{-0.070}$ & $0.986^{+0.063}_{-0.056}$ & $0.422^{+0.057}_{-0.059}$ \\
all systematics, $\spara^{\rm max}=180$\mpc, $z_{\rm mod}=0.82$ & $1.001^{+0.072}_{-0.072}$ & $1.006^{+0.062}_{-0.060}$ & $0.454^{+0.068}_{-0.067}$ \\
all systematics, $\spara^{\rm max}=180$\mpc, $z_{\rm mod}=0.83$ & $0.999^{+0.073}_{-0.076}$ & $1.001^{+0.063}_{-0.068}$ & $0.449^{+0.065}_{-0.066}$ \\
all systematics, $\spara^{\rm max}=180$\mpc, $z_{\rm mod}=0.84$ & $0.995^{+0.073}_{-0.072}$ & $0.998^{+0.063}_{-0.065}$ & $0.445^{+0.059}_{-0.061}$ \\

all systematics, $\spara^{\rm max}=190$\mpc, $z_{\rm mod}=0.82$ & $1.002^{+0.071}_{-0.073}$ & $1.007^{+0.066}_{-0.057}$ & $0.455^{+0.068}_{-0.066}$ \\
all systematics, $\spara^{\rm max}=190$\mpc, $z_{\rm mod}=0.84$ & $1.000^{+0.072}_{-0.073}$ & $1.001^{+0.062}_{-0.063}$ & $0.447^{+0.065}_{-0.069}$ \\

all systematics, $\spara^{\rm max}=200$\mpc, $z_{\rm mod}=0.82$ & $1.003^{+0.069}_{-0.072}$ & $1.009^{+0.062}_{-0.056}$ & $0.457^{+0.079}_{-0.074}$ \\
all systematics, $\spara^{\rm max}=200$\mpc, $z_{\rm mod}=0.83$ & $1.003^{+0.069}_{-0.071}$ & $1.008^{+0.061}_{-0.056}$ & $0.453^{+0.071}_{-0.071}$ \\
all systematics, $\spara^{\rm max}=200$\mpc, $z_{\rm mod}=0.84$ & $0.996^{+0.073}_{-0.063}$ & $1.002^{+0.061}_{-0.062}$ & $0.452^{+0.066}_{-0.065}$ \\
all systematics, $\spara^{\rm max}=200$\mpc, $z_{\rm mod}=0.87$ & $0.997^{+0.071}_{-0.069}$ & $0.996^{+0.062}_{-0.062}$ & $0.438^{+0.056}_{-0.058}$ \\
all systematics, no cut & $1.018^{+0.086}_{-0.082}$ & $1.011^{+0.060}_{-0.061}$ &  $0.436^{+0.110}_{-0.109}$ \\ 
all systematics, $+1/2$bins & $1.002^{+0.069}_{-0.070}$ & $1.008^{+0.064}_{-0.071}$ & $0.459^{+0.068}_{-0.073}$ \\
\hline 
\hline
RSD+BAO & \\
\hline
& Standard 2PCF \\
\hline
no systematics & $1.005^{+0.072}_{-0.073}$ & $1.012^{+0.050}_{-0.052}$ & $0.459^{+0.061}_{-0.059}$ \\ 
all systematics & $0.979^{+0.080}_{-0.083}$ & $0.969^{+0.062}_{-0.065}$ & $0.418^{+0.062}_{-0.058}$ \\ 
\hline
& Modified 2PCF \\
\hline
no systematics &  $0.996^{+0.067}_{-0.069}$ & $1.009^{+0.046}_{-0.045}$ & $0.462^{+0.064}_{-0.065}$ \\ 
all systematics & $0.997^{+0.068}_{-0.069}$ & $1.003^{+0.052}_{-0.053}$ & $0.455^{+0.066}_{-0.065}$ \\ 
all systematics, $+1/2$bins & $0.997^{+0.070}_{-0.068}$ & $1.004^{+0.054}_{-0.052}$ &  $0.460^{+0.071}_{-0.073}$ \\
\hline
\end{tabular} 
\caption{\label{singleEZmockfit} {\bf Results of} RSD and BAO+RSD fits on 1000 EZmocks. 
We present the median and the 0.16 and 0.84 quantiles of the distribution of the best-fit values. Except for the first measurement, we use 'shuffled-z' EZmocks. 
}
\end{table*}
\renewcommand{\arraystretch}{1.0}

\section{Results}  \label{sec:results}
In this Section we present the results and tests made on the eBOSS ELG  data sample. We perform RSD and combined RSD+isotropic BAO measurements. All results are reported in Table \ref{datafit}. 

Following \cite{demattia20a}, we decided to limit the redshift range for the RSD fit to $0.7<z<1.1$ due to the higher variations of the radial selection function with depth in the $0.6<z<0.7$ interval. The posteriors become also more stable with this restricted redshift range. Limiting the RSD fit to $0.7<z<1.1$ moves the effective redshift of the combined sample from 0.845 to 0.857 (Table \ref{ELG_stats}). As we still keep the full range for the BAO part of the joined fit, we chose to fix the effective redshift to $z_{\rm eff}=0.85$ for the combined RSD+BAO measurements. Indeed, as argued in \cite{demattia20a}, changing the effective redshift from 0.845 to 0.857 induces shifts in the cosmological parameter measurements of  0.3$\%$ for $f\sigma_8$, 0.7$\%$ for $D_H/r_{\rm drag}$ and 1.1$\%$ for $D_M/r_{d\rm rag}$, which are small compared to the statistical uncertainty.

\begin{figure}
\includegraphics[width=0.95\columnwidth]{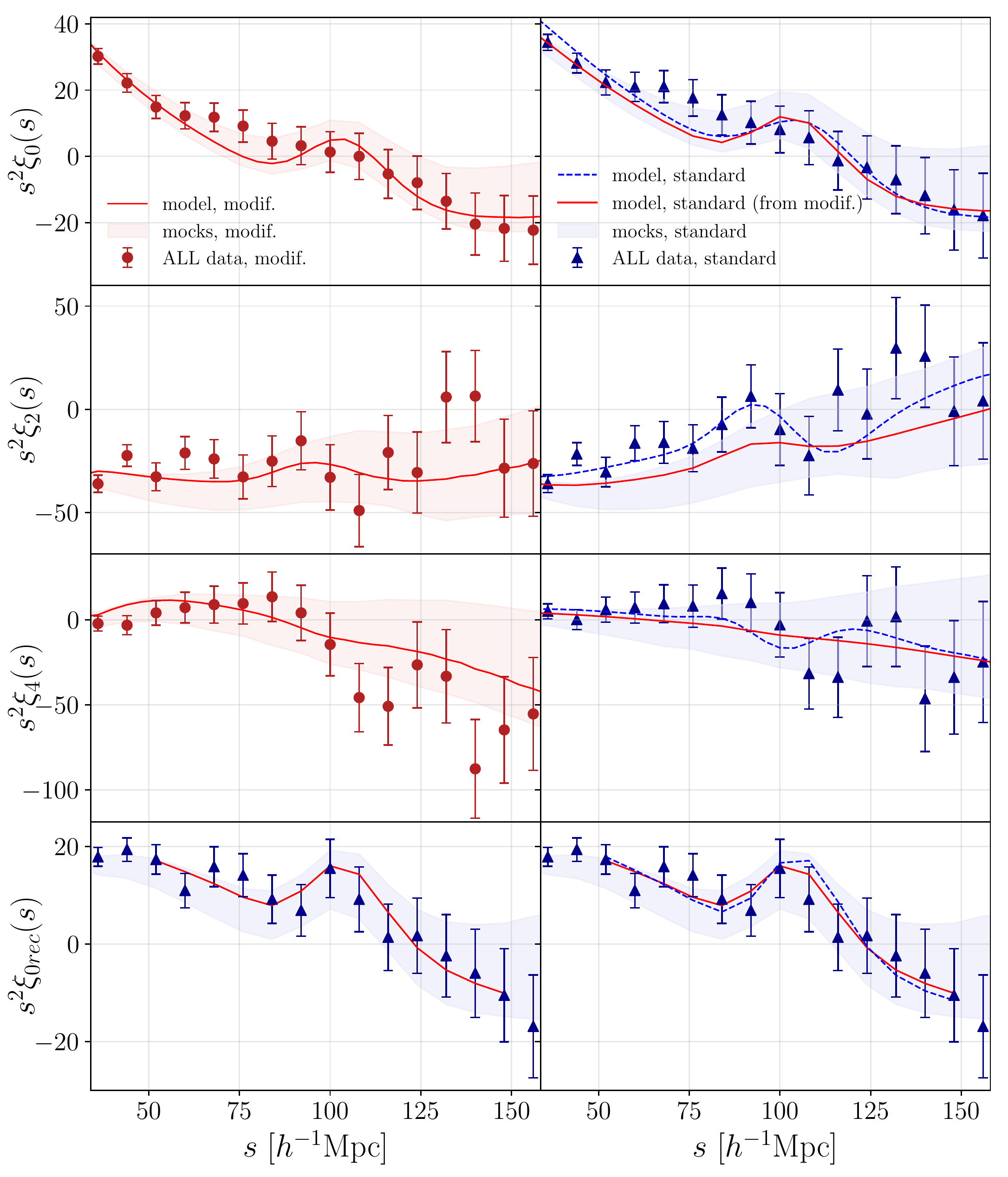}
\caption{\label{fig:bestfitsdata} 2PCF multipoles from eBOSS ELG data compared to CLPT-GS models. Left: pre-reconstruction mulipoles from the modified 2PCF of Equation \ref{eq:mod2pcffinal} with baseline parameter values, and post-reconstruction monopole from the standard 2PCF. The modifed 2PCF model (in red) is that from the RSD+BAO fit to the four multipoles in the left panels. Right: multipoles of the standard 2PCF compared to the standard 2PCF model with parameters from the RSD+BAO fit to the multipoles in the right panels (in blue) and in the left panels (in red). The bands are one sigma dispersions of the EZmocks for the modified (red) and standard (blue) 2PCF.} 
\end{figure}

\begin{figure}
\centering
\includegraphics[width=0.95\columnwidth]{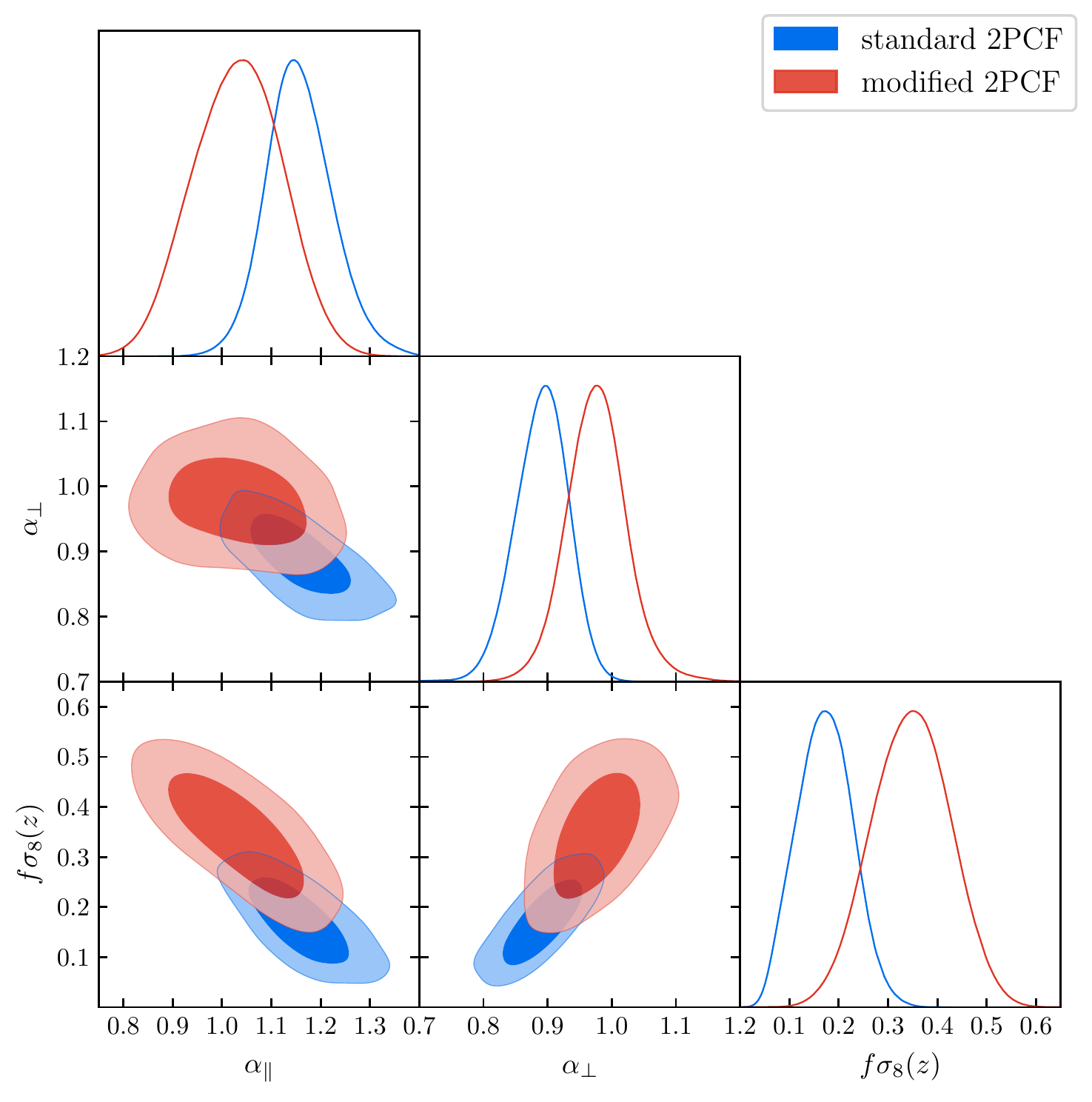}
\caption{\label{fig:posteriorsRSDBAO} Posteriors of the 
RSD+BAO fits to standard (in blue) and modified (in red) 2PCF multipoles as measured from the eBOSS ELG sample.}
\end{figure}

Results of RSD+BAO fits to the combined data sample are presented in Figure \ref{fig:bestfitsdata}, which compares data and best-fit model predictions for the post-reconstruction monopole and the pre-reconstructed 2PCF multipoles.
The right panel corresponds to results obtained with the standard 2PCF. While both monopole best-fits provide reasonable BAO peak positions, the quadrupole best-fit displays an unphysical 'BAO peak' at $s \sim 90\Mpch$, driven by a bump in the data, likely due to remaining angular systematics, which as a consequence biases the AP parameters. The degeneracy between the AP parameters and the growth rate observed in the posteriors, presented in Figure \ref{fig:posteriorsRSDBAO} (blue contours), can explain the low value measured for $f\sigma_8$. The fact that the model provides a good fit to all multipoles, including the quadrupole, explains the low $\chi^2$ obtained with the standard 2PCF, see Figure \ref{fig:EZmockschi2}.

Pure RSD fits on the eBOSS ELG sample with the standard 2PCF give results far away from what is expected from EZmocks, for the combined sample and separate caps. Compared to values measured in data ('baseline' of RSD Standard 2PCF), RSD fits to EZmocks with systematics using the standard 2PCF provide a larger value of $\alpha_{\parallel}$ in 33/1000 cases and the same fraction provides a smaller value of $\alpha_{\perp}$. However we observe no mock with a value of $f\sigma_8$ smaller than that in data and only a few mocks with a value around 35$\%$ larger. We interpret those unlikely results as due to the remaining angular systematics present in the data and to the low significance BAO detection in the eBOSS ELG sample presented in \cite{raichoor20a}. Changing the redshift range to $z_{\rm min}=0.6$ gives even more extreme results, with 14/1000 and 13/1000 mocks showing larger values of  $\alpha_{\parallel}$ and lower values of $\alpha_{\perp}$ than in data, respectively.
Adding the isotropic BAO to the fit ('baseline' of RSD+BAO Standard 2PCF) brings only slight changes to the previous results: 25/1000 mocks have a larger value than that measured for $\alpha_{\parallel}$, 132/1000 have a smaller value for $\alpha_{\perp}$ and 2/1000 mocks have a smaller value for $f\sigma_8$. The data measurements are still 
far from expected in the mocks. 

To mitigate the remaining angular systematics in the data sample, we fit the modified 2PCF from Equation \ref{eq:mod2pcffinal} with the same baseline parameter values as for the EZmocks, i.e. $z_{\rm mod}=0.83$ and $s_\parallel^{\rm max}=190\Mpch$, for which we observed that the systematic effects injected in the mocks were optimally reduced.

Cosmological parameter measurements for pure RSD fits with the modified 2PCF ('baseline' of RSD Modified 2PCF) are significantly different from those with the standard 2PCF: the value of $\alpha_{\parallel}$ decreases by 17.8$\%$ and those of $\alpha_{\perp}$ and $f\sigma_8$ increase by 12.6$\%$ and 146.5$\%$, respectively. Now 293/1000 mocks have a smaller value of $\alpha_{\parallel}$, 283/1000 a smaller value of $\alpha_{\perp}$ and there are 135/1000 mocks with a smaller value of $f\sigma_8$. Overall the new measurements are all within one sigma from the median of the fits to 'shuffled-z' EZmocks with systematics using the modified 2PCF. Larger differences between fits to data with the standard and modified 2PCFs are 
observed 
in 36/1000 and 80/1000 mocks for $\alpha_{\parallel}$ and $\alpha_{\perp}$. However no fit on mocks exhibits a difference as large as that in data for $f\sigma_8$.

Adding the post-reconstruction monopole of the standard 2PCF to the pre-reconstruction multipoles of the modifed 2PCF for a joined RSD+BAO fit ('baseline' of RSD+BAO Modified 2PCF) changes the previous results of 
pure RSD fits, increasing the value of $\alpha_{\parallel}$ by 8.2\%, that of $\alpha_{\perp}$ by 2.3\% and decreasing the value of $f\sigma_8$ by 8.9\%. There are 285/1000 mocks with a higher value of $\alpha_{\parallel}$, 388/1000 with a lower value of $\alpha_{\perp}$ and 61/1000 mocks with a lower value of $f\sigma_8$. In terms of the BAO isotropic shift derived from RSD fits using the modified 2PCF, adding the post-reconstruction monopole increases the value of $\alpha_{\rm iso}$ from $0.949$ ('baseline' of RSD Modified 2PCF) to $0.995$ ('baseline' of RSD+BAO Modified 2PCF) which is more consistent with the value measured by \citet{raichoor20a}. 
Compared with the results from BAO+RSD fits using the standard 2PCF, the value of $\alpha_{\parallel}$ decreases by 10.4$\%$, while those of $\alpha_{\perp}$ and $f\sigma_8$ increase by 9.4$\%$ and 103.5\%, respectively ('baseline' of RSD+BAO Modified vs Standard 2PCF). The differences in measured parameter values between 
fits using the standard or modified 2PCF are more frequent on RSD+BAO fits 
to EZmocks with systematics than for pure RSD fits: 154/1000 mocks have a larger shift than the observed one for $\alpha_{\parallel}$ and 139/1000 for $\alpha_{\perp}$ instead of 
36/1000 and 80/1000, respectively, for pure RSD fits as stated 
above. For $f\sigma_8$ there is still no mock for which such a difference is observed. We conclude that, as already observed on mocks, the modified 2PCF, being less prone to systematics, 
provides a more reliable estimator to derive cosmological measurements from data and that adding BAO regularizes the measurements.

The left panel of Figure \ref{fig:bestfitsdata} shows the pre-reconstruction multipoles and the post-reconstruction monopole of the modified 2PCF used for the RSD+BAO fits along with predictions from the best-fit model. The agreement between the best-fit model and the measured multipoles is good and the excess of clustering in the quadrupole at intermediate scale is significantly reduced in data, no longer driving the fit. On the right panel we show the predictions from the standard 2PCF model using best-fit values from the RSD+BAO fit with the modified 2PCF (in red on the graph). The model agrees quite well with the measured standard 2PCF multipoles, except 
at intermediate scales 
for the quadrupole, which are contaminated by systematics; we also note a better agreement for the lower $s$ bins for the monopole. 
The posteriors of the modified 2PCF RSD+BAO fit are presented in Figure~\ref{fig:posteriorsRSDBAO} (red contours).  As discussed above, removing angular modes with the modified 2PCF leads to different cosmological parameter 
estimates than with the standard 2PCF, though with similar degeneracies. We also note that due to information loss with the modified 2PCF, the posteriors are slightly wider than in the standard case. 

We now test the robustness of the  results 
from the above analysis with the modified 2PCF. Parameters of the latter were varied, removing the cut in the correction terms (i.e. using Equation~\ref{eq:mod2pcf}), 
and varying $z_{\rm mod}$ and $\spara^{\rm max}$ values, since, as stated in Section~\ref{sec:mitigatesys}, those are the most sensitive parameters. 
As for 
EZmocks, we vary $z_{\rm mod}$ and $\spara^{\rm max}$ values
in the ranges $\Delta z_{\rm mod}\in\{0,\pm0.01\}$ and $\Delta \spara^{\rm max}\in\{0,\pm10\}\Mpch$ around their baseline values.
We note that within the explored region, 
deviations in data measurements from the baseline results are in agreement with 
expectations from the mocks. Indeed staying on the diagonal defined by the crosses in Figure~\ref{fig:chizRPfig} gives small shifts with respect to 
baseline measurements and for most of the 
tested ($z_{\rm mod}$, $\spara^{\rm max}$) values, the deviations increase 
in accordance with the $\chi_{\rm mod}^2(z_{\rm mod}, \spara^{\rm max})$ value from the mocks. 
In agreement with the mocks, the largest 
deviations 
are observed for 
$z_{\rm mod}=0.84$ and $\spara^{\rm max}=180$. Shifts 
with respect to our baseline results in the pure RSD case 
amount to 8.0\%, 1.9\% and 20.7\% in $\alpha_{\parallel}$, $\alpha_{\perp}$ and $f\sigma_8$ respectively. Shifts are slightly smaller in the RSD+BAO case: 4.2\%, 1.8\% and 17.2\%. In the RSD case, such deviations are 
consistent with mocks at the 3$\sigma$ level for $\apara$ and 
at the 2$\sigma$ level for $\alpha_{\perp}$, 
but no mock shows a difference as large as for data for $f\sigma_8$. As those parameter values are not optimal for our analysis (see Figure~\ref{fig:chizRPfig}) large shifts are not surprising. Moreover, we know that our data sample suffers from systematic effects that are more complex than those introduced in the mocks, as observed when using the standard 2PCF (see Figure~\ref{fig:datacorr}). Nevertheless, we adopt a conservative approach and 
add the above shifts, i.e. 4.2\%, 1.8\% and 17.2\% in $\alpha_{\parallel}$, $\alpha_{\perp}$ and $f\sigma_8$, to our systematic budget to account for residual, uncorrected systematics in data. This error also includes the uncertainty due to the sensitivity of our results to the modified 2PCF free parameters.

When moving the $s$-bin centres by half a bin width (i.e. 4$\Mpch$), we observe large changes especially in $\alpha_{\parallel}$. The shifts for RSD+BAO fits are 5.7\%, 0.2\%, 1.7\% in $\alpha_{\parallel}$, $\alpha_{\perp}$ and $f\sigma_8$, respectively. Larger shifts are observed in 124/1000, 788/1000 and 766/1000 mocks in $\alpha_{\parallel}$, $\alpha_{\perp}$ and $f\sigma_8$ respectively. The observed shifts in data are therefore compatible with statistical fluctuations.

The measurements are stable when using the covariance matrix from 'shuffled-z' EZmocks without systematics: in the RSD+BAO case, we observe shifts 
of 0.6\%, 0.2\%, 1.7\% in $\alpha_{\parallel}$, $\alpha_{\perp}$ and $f\sigma_8$, compatible with statistical fluctuations. They remain stable also when we remove the $w_{\rm noz}$ weights when computing the correlation function: we observe small shifts of 0.3\% in $\alpha_{\parallel}$, 0.4\% in $\alpha_{\perp}$ and 1.7\% in $f\sigma_8$.
We finally checked the impact of changing the BOSS fiducial cosmology (Equation \ref{eq:cosmoBOSS}) to the OR one (Equation \ref{eq:cosmoOU}). Compared to our baseline results in the pure RSD case, we see deviations of 2.1\% in $\alpha_{\parallel}$, 0.3\% in $\alpha_{\perp}$ and 2.6\% in $f\sigma_8$. Those deviations are compatible with statistical fluctuations and considering the large systematic uncertainty already included for data instabilities, we do not 
add an extra systematic error.

Taking into account all systematic uncertainties from Table \ref{tab:budget} and adding them in quadrature to statistical errors, we quote our final measurements from the joined RSD+BAO fit with multipoles of the modifed 2PCF at the effective redshift $z_{\rm eff}=0.85$:
\begin{equation}\label{eq:res_alphas}
\alpha_{\parallel}=1.034^{+0.105}_{-0.111}\,,\quad \alpha_{\perp}=0.976 ^{+0.051}_{-0.051}\,,\quad f\sigma_8=0.348^{+0.103}_{-0.104}.
\end{equation}
The linear bias of our combined data sample for a $\sigma_8$ fixed at our fiducial cosmology (Equation \ref{eq:cosmoBOSS}) is measured to be 
$b_1=1.52^{+0.16}_{-0.14}$, where quoted errors are statistical only.

Converting the AP parameters into 
Hubble and comoving angular distances using Equations \ref{eq:alphasAP}, we finally have:
\begin{equation}\hspace*{2cm}
\begin{split}
D_H(z_{\rm eff})/r_{\rm drag}&=19.1^{+1.9}_{-2.1}\\ 
D_M(z_{\rm eff})/r_{\rm drag}&=19.9\pm1.0 \\
f\sigma_8(z_{\rm eff})&=0.35\pm0.10
\end{split}
\label{eq:cosmoMD}
\end{equation}

Those values are in agreement within less than one sigma with the values measured in Fourier space as reported in \citet{demattia20a}.
This allows to combine our two measurements into a consensus one for the eBOSS ELG sample, as presented in \citet{demattia20a}:
\begin{equation}\hspace*{2cm}
\begin{split}
D_H(z_{\rm eff})/r_{\rm drag}&=19.6^{+2.2}_{-2.1}\\ 
D_M(z_{\rm eff})/r_{\rm drag}&=19.5\pm1.0 \\
f\sigma_8(z_{\rm eff})&=0.315\pm0.095
\end{split}
\label{eq:cosmoMD}
\end{equation}
These results are compatible with a $\Lambda$CDM model using a Planck cosmology.

\renewcommand{\arraystretch}{1.0}
\begin{table*}
\centering
\begin{tabular}{|l@{\hspace{2cm}}|c@{\qquad}|c@{\qquad}|c@{\qquad}|}
\hline 
 & $\alpha_{\parallel}$ & $\alpha_{\perp}$ & $f\sigma_8$ \\ 
\hline 
\hline
RSD & \\
\hline
& Standard 2PCF \\
\hline
baseline & $1.163^{+0.087}_{-0.083}$ ($1.159$) & $0.847^{+0.071}_{-0.082}$ ($0.855$) & $0.155^{+0.069}_{-0.060}$ ($0.074$)\\
$z_{\rm min}=0.6$ & $1.212^{+0.086}_{-0.088}$ ($1.188$) & $0.801^{+0.096}_{-0.109}$ ($0.847$) & $0.100^{+0.079}_{-0.069}$ ($0.061$)\\
\hline
& Modified 2PCF \\
\hline
\bf{baseline (\boldmath$\spara^{\rm max}=190$\mpc, \boldmath$z_{\rm mod}=0.83$)} & \boldmath$0.956^{+0.125}_{-0.109}$ ($0.863$) & \boldmath$0.954^{+0.046}_{-0.050}$ ($0.950$) & \boldmath$0.382^{+0.078}_{-0.094}$ ($0.424$)\\
no sys. cov & $0.983^{+0.132}_{-0.141}$ ($0.854$) & $0.965^{+0.050}_{-0.050}$ ($0.954$) & $0.373^{+0.083}_{-0.100}$ ($0.429$)\\
no $w_{\rm noz}$ & $0.949^{+0.13}_{-0.107}$ ($0.862$) & $0.956^{+0.055}_{-0.055}$ ($0.951$) & $0.385^{+0.081}_{-0.083}$ ($0.423$)\\
$+1/2$bins & $0.864^{+0.154}_{-0.088}$ ($0.813$) & $0.946^{+0.057}_{-0.054}$ ($0.942$) & $0.394^{+0.077}_{-0.095}$ ($0.405$)\\
OR cosmology (rescaled) & $0.976^{+0.113}_{-0.102}$ ($0.907$) & $0.951^{+0.050}_{-0.045}$ ($0.954$) & $0.372^{+0.080}_{-0.093}$ ($0.401$)\\
$z_{\rm min}=0.6$ & $1.018^{+0.121}_{-0.121}$ ($0.929$) & $0.935^{+0.039}_{-0.045}$ ($0.942$) & $0.323^{+0.081}_{-0.090}$ ($0.366$)\\
$\spara^{\rm max}=180$\mpc, $z_{\rm mod}=0.82$ & $0.968^{+0.124}_{-0.115}$ ($0.862$) & $0.948^{+0.050}_{-0.051}$ ($0.948$) & $0.368^{+0.079}_{-0.088}$ ($0.421$)\\
$\spara^{\rm max}=180$\mpc, $z_{\rm mod}=0.83$ & $0.993^{+0.121}_{-0.131}$ ($0.862$) & $0.945^{+0.049}_{-0.053}$ ($0.935$) & $0.348^{+0.084}_{-0.099}$ ($0.404$)\\
$\spara^{\rm max}=180$\mpc, $z_{\rm mod}=0.84$ & $1.029^{+0.104}_{-0.141}$ ($0.875$) & $0.938^{+0.050}_{-0.051}$ ($0.931$) & $0.311^{+0.092}_{-0.087}$ ($0.382$)\\
$\spara^{\rm max}=190$\mpc, $z_{\rm mod}=0.82$ & $0.925^{+0.139}_{-0.094}$ ($0.855$) & $0.952^{+0.055}_{-0.052}$ ($0.951$) & $0.404^{+0.077}_{-0.088}$ ($0.438$)\\
$\spara^{\rm max}=190$\mpc, $z_{\rm mod}=0.84$ & $0.988^{+0.118}_{-0.125}$ ($0.869$) & $0.946^{+0.047}_{-0.052}$ ($0.943$) & $0.350^{+0.085}_{-0.093}$ ($0.404$)\\
$\spara^{\rm max}=200$\mpc, $z_{\rm mod}=0.82$ & $0.906^{+0.138}_{-0.083}$ ($0.847$) & $0.970^{+0.062}_{-0.053}$ ($0.956$) & $0.444^{+0.075}_{-0.082}$ ($0.458$)\\
$\spara^{\rm max}=200$\mpc, $z_{\rm mod}=0.83$ & $0.908^{+0.126}_{-0.085}$ ($0.852$) & $0.961^{+0.051}_{-0.051}$ ($0.955$) & $0.424^{+0.074}_{-0.081}$ ($0.446$)\\
$\spara^{\rm max}=200$\mpc, $z_{\rm mod}=0.84$ & $0.934^{+0.136}_{-0.093}$ ($0.880$) & $0.957^{+0.049}_{-0.055}$ ($0.953$) & $0.391^{+0.079}_{-0.085}$ ($0.226$)\\
no cut & $0.936^{+0.190}_{-0.108}$ ($0.84$) & $0.958^{+0.061}_{-0.061}$ ($0.958$) & $0.405^{+0.133}_{-0.196}$ ($0.458$)\\
\hline
& Separate caps \\
\hline
SGC, standard 2PCF & $1.100^{+0.090}_{-0.085}$ ($1.100$) & $0.946^{+0.077}_{-0.078}$ ($0.955$) & $0.236^{+0.082}_{-0.087}$ ($0.215$)\\
SGC, modified 2PCF & $1.041^{+0.093}_{-0.097}$ ($1.032$) & $1.026^{+0.118}_{-0.091}$ ($1.008$) & $0.378^{+0.102}_{-0.116}$ ($0.329$)\\
NGC, standard 2PCF & $1.196^{+0.113}_{-0.212}$ ($1.400$) & $0.759^{+0.085}_{-0.067}$ ($0.725$) & $0.147^{+0.095}_{-0.059}$ ($0.060$)\\
NGC, modified 2PCF & $0.875^{+0.377}_{-0.089}$ ($0.822$) & $0.932^{+0.371}_{-0.104}$ ($0.921$) & $0.463^{+0.095}_{-0.108}$ ($0.464$)\\
\hline 
\hline
RSD+BAO & \\
\hline
& Standard 2PCF \\
\hline
baseline & $1.154^{+0.071}_{-0.063}$ ($1.153$) & $0.892^{+0.040}_{-0.045}$ ($0.909$) & $0.171^{+0.058}_{-0.059}$ ($0.157$)\\
$z_{\rm min}=0.6$ & $1.198^{+0.060}_{-0.069}$ ($1.183$) & $0.846^{+0.046}_{-0.045}$ ($0.860$) & $0.109^{+0.064}_{-0.059}$ ($0.104$)\\
\hline
& Modified 2PCF \\
\hline
\bf{baseline (\boldmath$\spara^{\rm max}=190$\mpc, \boldmath$z_{\rm mod}=0.83$}) & \boldmath$1.034^{+0.091}_{-0.098}$ ($1.042$) & \boldmath$0.976^{+0.045}_{-0.045}$ ($0.978$) & \boldmath$0.348^{+0.082}_{-0.084}$ ($0.316$)\\
no sys. cov & $1.040^{+0.093}_{-0.112}$ ($1.050$) & $0.974^{+0.046}_{-0.043}$ ($0.978$) & $0.342^{+0.086}_{-0.091}$ ($0.308$)\\
no $w_{\rm noz}$ & $1.037^{+0.089}_{-0.097}$ ($1.044$) & $0.980^{+0.044}_{-0.044}$ ($0.981$) & $0.342^{+0.088}_{-0.083}$ ($0.314$)\\
$+1/2$bins & $0.975^{+0.120}_{-0.101}$ ($0.904$) & $0.978^{+0.055}_{-0.049}$ ($1.003$) & $0.354^{+0.094}_{-0.110}$ ($0.378$)\\
$z_{\rm min}=0.6$ & $1.082^{+0.083}_{-0.107}$ ($1.098$) & $0.950^{+0.035}_{-0.041}$ ($0.954$) & $0.299^{+0.080}_{-0.076}$ ($0.262$)\\
$\spara^{\rm max}=180$\mpc, $z_{\rm mod}=0.82$ & $1.047^{+0.085}_{-0.106}$ ($1.049$) & $0.971^{+0.043}_{-0.046}$ ($0.972$) & $0.333^{+0.085}_{-0.083}$ ($0.304$)\\
$\spara^{\rm max}=180$\mpc, $z_{\rm mod}=0.83$ & $1.057^{+0.082}_{-0.097}$ ($1.069$) & $0.966^{+0.045}_{-0.043}$ ($0.966$) & $0.317^{+0.082}_{-0.076}$ ($0.281$)\\
$\spara^{\rm max}=180$\mpc, $z_{\rm mod}=0.84$ & $1.077^{+0.077}_{-0.100}$ ($1.087$) & $0.958^{+0.042}_{-0.043}$ ($0.957$) & $0.288^{+0.081}_{-0.073}$ ($0.255$)\\
$\spara^{\rm max}=190$\mpc, $z_{\rm mod}=0.82$ & $1.027^{+0.087}_{-0.103}$ ($0.989$) & $0.978^{+0.050}_{-0.045}$ ($0.987$) & $0.361^{+0.084}_{-0.086}$ ($0.365$)\\
$\spara^{\rm max}=190$\mpc, $z_{\rm mod}=0.84$ & $1.058^{+0.082}_{-0.090}$ ($1.069$) & $0.968^{+0.044}_{-0.045}$ ($0.967$) & $0.319^{+0.080}_{-0.081}$ ($0.282$)\\
$\spara^{\rm max}=200$\mpc, $z_{\rm mod}=0.82$ & $0.988^{+0.096}_{-0.089}$ ($0.946$) & $0.991^{+0.051}_{-0.047}$ ($0.997$) & $0.398^{+0.079}_{-0.087}$ ($0.410$)\\
$\spara^{\rm max}=200$\mpc, $z_{\rm mod}=0.83$ & $1.009^{+0.093}_{-0.097}$ ($0.950$) & $0.988^{+0.050}_{-0.046}$ ($0.995$) & $0.383^{+0.079}_{-0.089}$ ($0.401$)\\
$\spara^{\rm max}=200$\mpc, $z_{\rm mod}=0.84$ & $1.028^{+0.088}_{-0.099}$ ($0.980$) & $0.979^{+0.044}_{-0.045}$ ($0.992$) & $0.351^{+0.083}_{-0.084}$ ($0.368$)\\
no cut & $1.036^{+0.107}_{-0.131}$ ($0.920$) & $0.972^{+0.059}_{-0.053}$ ($0.997$) & $0.339^{+0.145}_{-0.128}$ ($0.424$)\\
\hline
& Separate caps \\
\hline
SGC, standard 2PCF & $1.074^{+0.085}_{-0.084}$ ($1.069$) & $0.935^{+0.055}_{-0.052}$ ($0.938$) & $0.241^{+0.085}_{-0.080}$ ($0.225$)\\
SGC, modified 2PCF & $1.025^{+0.086}_{-0.093}$ ($1.023$) & $0.988^{+0.069}_{-0.062}$ ($0.979$) & $0.355^{+0.104}_{-0.104}$ ($0.314$)\\
NGC, standard 2PCF & $1.229^{+0.089}_{-0.097}$ ($1.204$) & $0.682^{+0.072}_{-0.035}$ ($0.874$) & $0.085^{+0.048}_{-0.019}$ ($0.163$)\\
NGC, modified 2PCF & $0.872^{+0.168}_{-0.078}$ ($0.824$) & $0.920^{+0.064}_{-0.072}$ ($0.906$) & $0.450^{+0.090}_{-0.106}$ ($0.458$)\\
\hline
\end{tabular} 
\caption{\label{datafit} RSD and BAO+RSD fits on the eBOSS ELG sample. We present the median and one sigma errors from the posterior distributions (as being the 0.16/0.84 quantiles from the distribution) and, in brackets, the best-fit value.}
\end{table*}
\renewcommand{\arraystretch}{1.0}

\begin{table}
\begin{tabular}{lccc}
\hline 
 & $\alpha_{\parallel}$ & $\alpha_{\perp}$ & $f\sigma_8$ \\ 
\hline 
From Nbody-mocks &  \\
\hline
CLPT modelling & 1.8\% & 1.4\% & 3.2\% \\
\hline
From EZmocks & \\
\hline
modelling RIC & 2.1\% & 1.2\% & 0.2\% \\
modified 2PCF & 0.5\% & 0.4\% & 2.7\% \\ 
\hline
From data & \\
\hline
uncorrected systematics & 4.2\% & 1.8\% & 17.2\% \\
\hline
\hline
Statistical uncertainties & $^{+0.091}_{-0.098}$ & $^{+0.045}_{-0.045}$ & $^{+0.082}_{-0.084}$ \\
Systematics uncertainties & $0.052$ & $0.025$ & $0.062$ \\
Total & $^{+0.105}_{-0.111}$ & $^{+0.051}_{-0.051}$ & $^{+0.103}_{-0.104}$ \\
\hline
\end{tabular} 
\caption{\label{tab:budget} Systematic error budget. The last row gives statistical and systematic errors added in quadrature.
}
\end{table}

\section{Conclusion}   \label{sec:conclusions}
We performed a pure RSD analysis and a joined RSD+BAO analysis in configuration space for the eBOSS DR16 Emission Line Galaxies sample described in \citet{raichoor20a}.
This sample is composed of 173,736 galaxies with a reliable redshift in the range $0.6<z<1.1$, covering an effective area of $\sim$730 deg$^2$ over the two NGC and SGC regions. The post-reconstruction BAO measurement in configuration space of this sample is analysed in \citet{raichoor20a}. The BAO and RSD measurements in Fourier space and a consensus of our results for the eBOSS ELG sample are presented in \citet{demattia20a}.

Our RSD fit is done on the $0.7<z<1.1$ data multipoles ($\ell=0,2,4$), using the CLPT-GS theoretical model.
As part of the eBOSS ELG mock challenge \citep{Alam2020}, we first demonstrate the validity of the CLPT-GS model in our fitting range using realistic ELG mocks.
Those are built from accurate N-body simulations, populated with a broad range of models describing ELG variety, and split into sets of 'non-blind' and 'blind' mocks.

A set of approximate mocks, the EZmocks \citep{zhao20a}, are used to estimate the covariance matrix and also to validate the analysis pipeline.
As for the data, those EZmocks have redshifts from randoms selected from the parent galaxy catalogue themselves, in order to properly reproduce the survey radial selection function.
However this choice leads to radial mode suppression, which we account for in the correlation function modelling with a correction based on the formalism developed in \citet{de-Mattia:2019aa}.
We validate and quantify the error budget coming from that correction using the EZmocks.

The eBOSS ELG data sample is affected by residual angular systematics, which need to be corrected for before proceeding to RSD fits, to avoid biasing our cosmological measurements.
To mitigate these angular systematics, we performed our RSD fits using a modified 2PCF estimator, which is computed consistently for the data, the EZmocks and the model, discarding the small scales where the accuracy of the CLPT-GS model is not demonstrated.
We carefully assessed the validity of that approach with a set of the EZmocks in which we injected data-like systematics.
We demonstrated the efficiency of our approach to remove angular systematics.

Once the validity of the RSD analysis and its error budget have been established,
we performed a similar analysis for the isotropic BAO measurement on the reconstructed monopole ($\ell = 0_{\rm rec}$).

Finally, we did a serie of tests on the RSD-only and RSD+BAO results from the ELG data sample. Due to the non-gaussianity of our results, the RSD+BAO joined fits are performed by combining their likelihoods.
Taking into account all systematic errors from our budget as well as statistical errors, we obtain our final measurements from the joined RSD+BAO fit to the modified 2PCF multipoles at the effective redshift $z_{\rm eff}=0.85$:
$\apara=1.034^{+0.105}_{-0.111}$,
$\aperp=0.976^{+0.051}_{-0.051}$, and
$\fsig=0.348^{+0.103}_{-0.104}$.
From this joined analysis we obtain $D_H(z_{\rm eff})/r_{\rm drag}=19.1^{+1.9}_{-2.1}$, $D_M(z_{\rm eff})/r_{\rm drag}=19.9\pm1.0$ and $f\sigma_8(z_{\rm eff})=0.35\pm0.10$.
These results are in agreement within less than 1$\sigma$ with those found by \citet{demattia20a} with a RSD+BAO analysis performed in Fourier space. We also present a consensus result between the two analyses, fully described in \citet{demattia20a}: $D_H(z_{\rm eff})/r_{\rm drag}=19.6^{+2.2}_{-2.1}$, $D_M(z_{\rm eff})/r_{\rm drag}=19.5\pm1.0$ and $f\sigma_8(z_{\rm eff})=0.315\pm0.095$, which are in agreement with $\Lambda$CDM predictions based on Planck parameters.

The presence of remaining angular systematics in the eBOSS ELG data led us to develop a specific analysis tool, the modified 2PCF estimator presented in this paper, that we consistently applied to the data, mocks and RSD model.
Such an approach, along with other developments based on the eBOSS data \citep{Kong2020,Mohammad2020,Rezaie:2020aa}, will pave the way for the analysis of the RSD and BAO in the next generation of surveys that massively rely on ELGs, such as DESI, \textit{Euclid}, PFS or \textit{WFIRST}.

\section*{Acknowledgements}
AT, AR and CZ acknowledge support from the SNF grant 200020\_175751.
AR, JPK acknowledge support from the ERC advanced grant LIDA.
AdM acknowledges support from the P2IO LabEx (ANR-10-LABX-0038) in the framework "Investissements d'Avenir" (ANR-11-IDEX-0003-01) managed by the Agence Nationale de la Recherche (ANR, France).
AJR is grateful for support from the Ohio State University Center for Cosmology and Particle Physics. SA is supported by the European Research Council (ERC) through the COSFORM Research Grant (\#670193). VGP acknowledges support from the European Union's Horizon 2020 research and innovation programme (ERC grant \#769130). EMM acknowledges support from the European Research Council (ERC) under the European Union's Horizon 2020 research and innovation programme (grant agreement No 693024). G.R. acknowledges support from the National Research Foundation of Korea (NRF) through Grants No. 2017R1E1A1A01077508 and No. 2020R1A2C1005655 funded by the Korean Ministry of Education, Science and Technology (MoEST), and from the faculty research fund of Sejong University. 
Authors acknowledge support from the ANR eBOSS project (ANR-16-CE31-0021) of the French National Research Agency.

Funding for the Sloan Digital Sky Survey IV has been provided by the Alfred P. Sloan Foundation, the U.S. Department of Energy Office of Science, and the Participating Institutions. SDSS-IV acknowledges
support and resources from the Center for High-Performance Computing at
the University of Utah. The SDSS web site is www.sdss.org.

SDSS-IV is managed by the Astrophysical Research Consortium for the 
Participating Institutions of the SDSS Collaboration including the 
Brazilian Participation Group, the Carnegie Institution for Science, Carnegie Mellon University, the Chilean Participation Group, the Ecole Polytechnique Federale de Lausanne (EPFL), the French Participation Group, Harvard-Smithsonian Center for Astrophysics, 
Instituto de Astrofisica de Canarias, The Johns Hopkins University, Kavli Institute for the Physics and Mathematics of the Universe (IPMU) \/ University of Tokyo, the Korean Participation Group, Lawrence Berkeley National Laboratory, 
Leibniz Institut f\"ur Astrophysik Potsdam (AIP),  
Max-Planck-Institut f\"ur Astronomie (MPIA Heidelberg), 
Max-Planck-Institut f\"ur Astrophysik (MPA Garching), 
Max-Planck-Institut f\"ur Extraterrestrische Physik (MPE), 
National Astronomical Observatories of China, New Mexico State University, 
New York University, University of Notre Dame, 
Observat\'ario Nacional / MCTI, The Ohio State University, 
Pennsylvania State University, Shanghai Astronomical Observatory, 
United Kingdom Participation Group,
Universidad Nacional Aut\'onoma de M\'exico, University of Arizona, 
University of Colorado Boulder, University of Oxford, University of Portsmouth, 
University of Utah, University of Virginia, University of Washington, University of Wisconsin, 
Vanderbilt University, and Yale University.

This research used resources of the National Energy Research Scientific Computing Center, a DOE Office of Science User Facility supported by the Office of Science of the U.S. Department of Energy under Contract No. DE-AC02-05CH11231.

\section*{Data Availability}
Correlation functions, covariance matrices, and resulting likelihoods for cosmological parameters are (will be made) available (after acceptance) via the SDSS Science Archive Server: https://sas.sdss.org/  (with the exact address tbd).




\bibliographystyle{mnras}
\bibliography{Ref} 



\appendix

\section{Modified Correlation - Details}\label{sec:appendix_modifcorr}
We provide in this appendix more context to the Equation \ref{eq:mod2pcf}. Starting from Equation 3.8 of \citet{Burden:2017aa}, the shuffled correlation function can be written for a normalised data density $\bar{n}(\chi)$ as:
\begin{align}
\xi^{\rm shuff}(\vect{r},\vect{r'})=&\langle\delta(\vect{r})\delta(\vect{r'})\rangle-2\big\langle\delta(\vect{r})\int\delta(\gamma, \chi')\bar{n}(\chi')d\chi'\big\rangle\\&+\big\langle\int\delta(\gamma, \chi')\bar{n}(\chi')d\chi'\int\delta(\gamma', \chi'')\bar{n}(\chi'')d\chi''\big\rangle
\end{align}
where $\delta$ is the density field,  and $\vect{r}$,  $\vect{r'}$ are the comoving positions, $\gamma$, $\gamma'$ are the corresponding angular positions and  
$\chi$ stands for line-of-sight positions. 

The first term corresponds to the standard 2PCF. Using the same approximation as in \citet{Burden:2017aa} (Equation 3.9) and doing the substitution $\chi'$ to $\Delta\chi=\chi'-\chi$, the second term becomes:
\begin{align}
\big\langle\delta(\vect{r})\int\delta(\gamma, \chi')\bar{n}(\chi')d\chi'\big\rangle&=\int\xi(\theta,\chi,\chi')\bar{n}(\chi')d\chi'\\&=\int\xi(\theta,\Delta\chi)\bar{n}(\chi+\Delta\chi)d\Delta\chi
\end{align}
writing $\theta=\gamma'-\gamma$. To be more flexible in the scales introduced in the correction, we further change $\bar{n}(\chi+\Delta\chi)$ to $\bar{n}(\chi_{\rm mod}+\Delta\chi/2)$ where $\chi_{\rm mod}=(\chi+\chi')/2$ is fixed (without changing the variable of integration). As already stated, we emphasize that such approximations have no impact on the validity of our analysis as we use the modified 2PCF as a new estimator applied consistently on data and model.

The third term corresponds to the angular correlation function $w(\theta)$. Using the same substitution as previously and the Limber approximation \citep{Limber:1953aa}, it becomes:
\begin{align}
w(\theta)&=\int\int\xi(\theta,\Delta\chi)\bar{n}(\chi)\bar{n}(\chi+\Delta\chi)d\chi d\Delta\chi\\&=\int\bar{n}^2(\chi)d\chi\int\xi(\theta,\Delta\chi)d\Delta\chi
\end{align}
Gathering all 
terms together we 
end up with the adopted modified 2PCF of Equation \ref{eq:mod2pcf}.


\bsp	
\label{lastpage}
\end{document}